\let\includefigures=\iftrue
%
\let\useblackboard=\iftrue
%
%
\newfam\black
\input harvmac
\noblackbox
\includefigures
\message{If you do not have epsf.tex (to include figures),}
\message{change the option at the top of the tex file.}
\input epsf
\def\figin{\epsfcheck\figin}\def\figins{\epsfcheck\figins}
\def\epsfcheck{\ifx\epsfbox\UnDeFiNeD
\message{(NO epsf.tex, FIGURES WILL BE IGNORED)}
\gdef\figin##1{\vskip2in}\gdef\figins##1{\hskip.5in}
\else\message{(FIGURES WILL BE INCLUDED)}%
\gdef\figin##1{##1}\gdef\figins##1{##1}\fi}
\def\DefWarn#1{}
\def\figinsert{\goodbreak\midinsert}
\def\ifig#1#2#3{\DefWarn#1\xdef#1{fig.~\the\figno}
\writedef{#1\leftbracket fig.\noexpand~\the\figno}%
\figinsert\figin{\centerline{#3}}\medskip\centerline{\vbox{
\baselineskip12pt\advance\hsize by -1truein
\noindent\footnotefont{\bf Fig.~\the\figno:} #2}}
\bigskip\endinsert\global\advance\figno by1}
\else
\def\ifig#1#2#3{\xdef#1{fig.~\the\figno}
\writedef{#1\leftbracket fig.\noexpand~\the\figno}%
\global\advance\figno by1}
\fi
%

\def\smallfig#1#2#3{\DefWarn#1\xdef#1{fig.~\the\figno}
\writedef{#1\leftbracket fig.\noexpand~\the\figno}%
\figinsert\figin{\centerline{#3}}\medskip\centerline{\vbox{
\baselineskip12pt\advance\hsize by -1truein
\noindent\footnotefont{\bf Fig.~\the\figno:} #2}}
\endinsert\global\advance\figno by1}

\useblackboard
\message{If you do not have msbm (blackboard bold) fonts,}
\message{change the option at the top of the tex file.}
\font\blackboard=msbm10 scaled \magstep1
\font\blackboards=msbm7
\font\blackboardss=msbm5
\textfont\black=\blackboard
\scriptfont\black=\blackboards
\scriptscriptfont\black=\blackboardss

\else

\fi
%



\def\boxit#1{\vbox{\hrule\hbox{\vrule\kern8pt
\vbox{\hbox{\kern8pt}\hbox{\vbox{#1}}\hbox{\kern8pt}}
\kern8pt\vrule}\hrule}}
\def\mathboxit#1{\vbox{\hrule\hbox{\vrule\kern8pt\vbox{\kern8pt
\hbox{$\displaystyle #1$}\kern8pt}\kern8pt\vrule}\hrule}}

\def\subsubsection#1{\bigskip\noindent
{\it #1}}

\def\yboxit#1#2{\vbox{\hrule height #1 \hbox{\vrule width #1
\vbox{#2}\vrule width #1 }\hrule height #1 }}
\def\fillbox#1{\hbox to #1{\vbox to #1{\vfil}\hfil}}
\def\ybox{{\lower 1.3pt \yboxit{0.4pt}{\fillbox{8pt}}\hskip-0.2pt}}
%
%


\def\bi{{\bar i}}
\def\jb{{\bar j}}
\def\bz{{\bar z}}

\def\tphi{{\tilde{\phi}}}

\def\l{\left}

\def\comments#1{}

\def\p{\partial}

\def\half{{1\over 2}}

\def\tr{{\rm tr\ }}
\def\Re{{\rm Re\hskip0.1em}}
\def\Im{{\rm Im\hskip0.1em}}

\def\ket#1{|#1\rangle}

\def\vev#1{\langle{#1}\rangle}

\def\CN{{\cal N}}
\def\CO{{\cal O}}


\def\II{\relax{I\kern-.10em I}}

\font\cmss=cmss10 \font\cmsss=cmss10 at 7pt
\def\IZ{\relax\ifmmode\mathchoice
{\hbox{\cmss Z\kern-.4em Z}}{\hbox{\cmss Z\kern-.4em Z}}
{\lower.9pt\hbox{\cmsss Z\kern-.4em Z}}
{\lower1.2pt\hbox{\cmsss Z\kern-.4em Z}}
\else{\cmss Z\kern-.4emZ}\fi}
\def\IR{\relax{\rm I\kern-.18em R}}
\def\IZ{\relax\ifmmode\mathchoice
{\hbox{\cmss Z\kern-.4em Z}}{\hbox{\cmss Z\kern-.4em Z}}
{\lower.9pt\hbox{\cmsss Z\kern-.4em Z}} {\lower1.2pt\hbox{\cmsss
Z\kern-.4em Z}}\else{\cmss Z\kern-.4em Z}\fi}
\def\IB{\relax{\rm I\kern-.18em B}}
\def\IC{{\relax\hbox{$\inbar\kern-.3em{\rm C}$}}}
\def\ID{\relax{\rm I\kern-.18em D}}
\def\IE{\relax{\rm I\kern-.18em E}}
\def\IF{\relax{\rm I\kern-.18em F}}
\def\IG{\relax\hbox{$\inbar\kern-.3em{\rm G}$}}
\def\IGa{\relax\hbox{${\rm I}\kern-.18em\Gamma$}}
\def\IH{\relax{\rm I\kern-.18em H}}
\def\II{\relax{\rm I\kern-.18em I}}
\def\IK{\relax{\rm I\kern-.18em K}}
\def\IP{\relax{\rm I\kern-.18em P}}

%

\def\jb{{\bar \jmath}}

\def\inbar{\,\vrule height1.5ex width.4pt depth0pt}

\def\p{\partial}

\font\cmss=cmss10 
\def\IR{\relax{\rm I\kern-.18em R}}

%


%

\def\lp10{\ell_p^{10}}
\def\lp11{\ell_p^{11}}
\def\R11{R_{11}}

\def\frac#1#2{{#1 \over #2}}



\def\l{\left}

\def\comments#1{}

\def\p{\partial}

\def\half{{1\over 2}}

\def\tr{{\rm tr\ }}
\def\Re{{\rm Re\hskip0.1em}}
\def\Im{{\rm Im\hskip0.1em}}

\def\ket#1{|#1\rangle}

\def\vev#1{\langle{#1}\rangle}

\def\CN{{\cal N}}
\def\CO{{\cal O}}



\def\qv{{\it q.v.}}
\def\M4{M_{Pl,4}}

\def\k11{\kappa_{11}}
\def\l11{\ell_{11}}
\def\tl11{\tilde{\ell}_{11}}

\def\m11{M_{11}}
\def\tm11{\tilde{M}_{11}}

\def\np{{\it Nucl. Phys.}}
\def\prl{{\it Phys. Rev. Lett.}}

\def\etal{{\it et.\ al.}}
\def\eg{{\it e.g.}}
\def\ie{{\it i.e.}}


%

\lref\bbs{
K.~Becker, M.~Becker and A.~Strominger,
``Five-branes, membranes and nonperturbative string theory,''
Nucl.\ Phys.\ B {\bf 456}, 130 (1995)
[arXiv:hep-th/9507158].
}

\lref\BrunnerJQ{
I.~Brunner, M.~R.~Douglas, A.~E.~Lawrence and C.~Romelsberger,
``D-branes on the quintic,''
JHEP {\bf 0008}, 015 (2000)
[arXiv:hep-th/9906200].
}
\lref\DouglasSW{
M.~R.~Douglas and G.~W.~Moore,
``D-branes, Quivers, and ALE Instantons,''
arXiv:hep-th/9603167.
}
\lref\DouglasDE{
M.~R.~Douglas, B.~R.~Greene and D.~R.~Morrison,
``Orbifold resolution by D-branes,''
Nucl.\ Phys.\ B {\bf 506}, 84 (1997)
[arXiv:hep-th/9704151].
}
\lref\DouglasGI{
M.~R.~Douglas,
``D-branes, categories and N = 1 supersymmetry,''
J.\ Math.\ Phys.\  {\bf 42}, 2818 (2001)
[arXiv:hep-th/0011017].
}

\lref\AspinwallDZ{
P.~S.~Aspinwall and M.~R.~Douglas,
``D-brane stability and monodromy,''
JHEP {\bf 0205}, 031 (2002)
[arXiv:hep-th/0110071].
}

\lref\AspinwallPU{
P.~S.~Aspinwall and A.~E.~Lawrence,
``Derived categories and zero-brane stability,''
JHEP {\bf 0108}, 004 (2001)
[arXiv:hep-th/0104147].
}

\lref\FayetYH{
P.~Fayet,
``Higgs Model And Supersymmetry,''
Nuovo Cim.\ A {\bf 31}, 626 (1976).
}

\lref\KachruVJ{
S.~Kachru and J.~McGreevy,
``Supersymmetric three-cycles and (super)symmetry breaking,''
Phys.\ Rev.\ D {\bf 61}, 026001 (2000)
[arXiv:hep-th/9908135].
}
\lref\JoyceTZ{
D.~Joyce,
``On counting special Lagrangian homology 3-spheres,''
Contemp.\ Math.\  {\bf 314}, 125 (2002)
[arXiv:hep-th/9907013].
}

\lref\wf{W. Fischler, H.P. Nilles, J. Polchinski, 
S. Raby, L. Susskind, ``Vanishing renormalization 
of the D term in supersymmetric U(1) theories,''
\prl\ {\bf 47} (1981) 757. }

\lref\DineXK{
M.~Dine, N.~Seiberg and E.~Witten,
``Fayet-Iliopoulos Terms In String Theory,''
Nucl.\ Phys.\ B {\bf 289}, 589 (1987).
}

\lref\ads{J.J. Atick, L.J. Dixon and A. Sen,
``String calculation of Fayet-Iliopolous D-terms
in arbitrary supersymmetric compactifications'',
\np\ {\bf B292}\ (1987) 109.}

\lref\DineGJ{
M.~Dine, I.~Ichinose and N.~Seiberg,
``F Terms And D Terms In String Theory,''
Nucl.\ Phys.\ B {\bf 293}, 253 (1987).
}

\lref\pop{E. Poppitz, ``On the one loop Fayet-Iliopoulos term in 
chiral four-dimensional type I orbifolds'', 
\np\ {\bf B542} (1999) 31 [arXiv:hep-th/9810010].}

\lref\bdl{
M.~Berkooz, M.~R.~Douglas and R.~G.~Leigh,
``Branes intersecting at angles,''
Nucl.\ Phys.\ B {\bf 480}, 265 (1996)
[arXiv:hep-th/9606139].
}

\lref\GreenDD{
M.~B.~Green, J.~A.~Harvey and G.~W.~Moore,
``I-brane inflow and anomalous couplings on D-branes,''
Class.\ Quant.\ Grav.\  {\bf 14}, 47 (1997)
[arXiv:hep-th/9605033].
}

\lref\DixonQV{
L.~J.~Dixon, D.~Friedan, E.~J.~Martinec and S.~H.~Shenker,
``The Conformal Field Theory Of Orbifolds,''
Nucl.\ Phys.\ B {\bf 282}, 13 (1987).
}

\lref\BinetruyHH{
P.~Binetruy, G.~Dvali, R.~Kallosh and A.~Van Proeyen,
``Fayet-Iliopoulos terms in supergravity and cosmology,''
arXiv:hep-th/0402046.
}
\lref\DvaliZH{
G.~Dvali, R.~Kallosh and A.~Van Proeyen,
``D-term strings,''
JHEP {\bf 0401}, 035 (2004)
[arXiv:hep-th/0312005].
}
\lref\MatsudaQT{
T.~Matsuda,
``Formation of cosmological brane defects,''
arXiv:hep-ph/0402232.
}
\lref\MatsudaBK{
T.~Matsuda,
``String production after angled brane inflation,''
arXiv:hep-ph/0403092.
}
\lref\HashimotoPU{
K.~Hashimoto and W.~Taylor,
``Strings between branes,''
JHEP {\bf 0310}, 040 (2003)
[arXiv:hep-th/0307297].
}

\lref\GiddingsYU{
S.~B.~Giddings, S.~Kachru and J.~Polchinski,
``Hierarchies from fluxes in string compactifications,''
Phys.\ Rev.\ D {\bf 66}, 106006 (2002)
[arXiv:hep-th/0105097].
}
\lref\KachruHE{
S.~Kachru, M.~B.~Schulz and S.~Trivedi,
``Moduli stabilization from fluxes in a simple IIB orientifold,''
JHEP {\bf 0310}, 007 (2003)
[arXiv:hep-th/0201028].
}
\lref\KachruAW{
S.~Kachru, R.~Kallosh, A.~Linde and S.~P.~Trivedi,
``De Sitter vacua in string theory,''
Phys.\ Rev.\ D {\bf 68}, 046005 (2003)
[arXiv:hep-th/0301240].
}
\lref\DenefDM{
F.~Denef, M.~R.~Douglas and B.~Florea,
``Building a better racetrack,''
arXiv:hep-th/0404257.
}
\lref\RobbinsHX{
D.~Robbins and S.~Sethi,
``A barren landscape,''
arXiv:hep-th/0405011.
}
\lref\vijayper{
V.~Balasubramanian and P.~Berglund,
``Stringy corrections to Kaehler potentials, SUSY breaking, and the
cosmological constant problem,''
arXiv:hep-th/0408054.
}

\lref\BinetruyXJ{
P.~Binetruy and G.~R.~Dvali,
``D-term inflation,''
Phys.\ Lett.\ B {\bf 388}, 241 (1996)
[arXiv:hep-ph/9606342].
}

\lref\joebook{
J.~Polchinski, {\it Superstring Theory}, vols. I-II, Cambridge Univ. Press.}
\lref\tatatheta{D.~Mumford, {\it Tata Lectures on Theta, vol. I}, Birkh\"auser (1983).}

\lref\LMtwo{
A.~Lawrence and J.~McGreevy,
``Remarks on branes, fluxes, and soft SUSY breaking,''
Proceedings of 3rd International Symposium on Quantum Theory and Symmetries (QTS3), Cincinnati, Ohio, 10-14 Sep 2003, 
arXiv:hep-th/0401233.
}

\lref\LMone{
A.~Lawrence and J.~McGreevy,
``Local string models of soft supersymmetry breaking,''
JHEP {\bf 0406}, 007 (2004)
[arXiv:hep-th/0401034].
}

\lref\AlvarezGaumeIG{
L.~Alvarez-Gaume and E.~Witten,
``Gravitational Anomalies,''
Nucl.\ Phys.\ B {\bf 234}, 269 (1984).
}

\lref\GreenSG{
M.~B.~Green and J.~H.~Schwarz,
``Anomaly Cancellation In Supersymmetric D=10 Gauge Theory And Superstring
Theory,''
Phys.\ Lett.\ B {\bf 149}, 117 (1984).
}

\lref\PeetES{
A.~W.~Peet,
``The Bekenstein formula and string theory (N-brane theory),''
Class.\ Quant.\ Grav.\  {\bf 15}, 3291 (1998)
[arXiv:hep-th/9712253].
}

\lref\DvaliPA{G.~R.~Dvali and S.~H.~H.~Tye,
``Brane inflation,'' Phys.\ Lett.\ B {\bf 450}, 72 (1999) [arXiv:hep-ph/9812483].}
\lref\AlexanderKS{S.~H.~S.~Alexander,``Inflation from D - anti-D brane annihilation,''
Phys.\ Rev.\ D {\bf 65}, 023507 (2002) [arXiv:hep-th/0105032].}
\lref\BurgessFX{C.~P.~Burgess, M.~Majumdar, D.~Nolte, F.~Quevedo, G.~Rajesh and R.~J.~Zhang,
``The inflationary brane-antibrane universe,''
JHEP {\bf 0107}, 047 (2001)[arXiv:hep-th/0105204].
}
\lref\DvaliFW{
G.~R.~Dvali, Q.~Shafi and S.~Solganik,
``D-brane inflation,''
arXiv:hep-th/0105203.
}
\lref\KachruSX{S.~Kachru, R.~Kallosh, A.~Linde, J.~Maldacena, L.~McAllister and S.~P.~Trivedi,
``Towards inflation in string theory,'' JCAP {\bf 0310}, 013 (2003)
[arXiv:hep-th/0308055].}
\lref\SarangiYT{
S.~Sarangi and S.~H.~H.~Tye,
``Cosmic string production towards the end of brane inflation,''
Phys.\ Lett.\ B {\bf 536}, 185 (2002)
[arXiv:hep-th/0204074].
}
\lref\GarciaBellidoKY{
J.~Garcia-Bellido, R.~Rabadan and F.~Zamora,
``Inflationary scenarios from branes at angles,''
JHEP {\bf 0201}, 036 (2002)
[arXiv:hep-th/0112147].
}

\lref\JonesDA{
N.~T.~Jones, H.~Stoica and S.~H.~H.~Tye,
``The production, spectrum and evolution of cosmic strings in brane inflation,''
Phys.\ Lett.\ B {\bf 563}, 6 (2003)
[arXiv:hep-th/0303269].
}

\lref\CopelandBJ{E.~J.~Copeland, R.~C.~Myers and J.~Polchinski,
``Cosmic F- and D-strings,''
JHEP {\bf 0406}, 013 (2004) [arXiv:hep-th/0312067].}

\lref\CopelandVG{
E.~J.~Copeland, A.~R.~Liddle, D.~H.~Lyth, E.~D.~Stewart and D.~Wands,
``False vacuum inflation with Einstein gravity,''
Phys.\ Rev.\ D {\bf 49}, 6410 (1994)
[arXiv:astro-ph/9401011].
}
\lref\StewartTS{
E.~D.~Stewart,
``Inflation, supergravity and superstrings,''
Phys.\ Rev.\ D {\bf 51}, 6847 (1995)
[arXiv:hep-ph/9405389].
}

\lref\ArkaniHamedMZ{
N.~Arkani-Hamed, H.~C.~Cheng, P.~Creminelli and L.~Randall,
``Pseudonatural inflation,''
JCAP {\bf 0307}, 003 (2003)
[arXiv:hep-th/0302034].}

\lref\bain{
P.~Bain and M.~Berg,
``Effective action of matter fields in four-dimensional string  orientifolds,''
JHEP {\bf 0004}, 013 (2000)
[arXiv:hep-th/0003185].
}

\lref\ibanez{
L.~E.~Ibanez, R.~Rabadan and A.~M.~Uranga,
``Sigma-model anomalies in compact D = 4, N = 1 type IIB orientifolds and
Fayet-Iliopoulos terms,''
Nucl.\ Phys.\ B {\bf 576}, 285 (2000)
[arXiv:hep-th/9905098].
}

\lref\DasguptaDW{
K.~Dasgupta, J.~P.~Hsu, R.~Kallosh, A.~Linde and M.~Zagermann,
``D3/D7 brane inflation and semilocal strings,''
JHEP {\bf 0408}, 030 (2004)
[arXiv:hep-th/0405247].
}

\lref\HalyoUU{
E.~Halyo,
``Cosmic D-term strings as wrapped D3 branes,''
JHEP {\bf 0403}, 047 (2004)
[arXiv:hep-th/0312268].
}

\lref\HalyoZJ{
E.~Halyo,
``D-brane inflation on conifolds,''
arXiv:hep-th/0402155.
}
\lref\HalyoPP{
E.~Halyo,
``Hybrid inflation from supergravity D-terms,''
Phys.\ Lett.\ B {\bf 387}, 43 (1996)
[arXiv:hep-ph/9606423].
} 

\lref\LeblondUC{
L.~Leblond and S.~H.~H.~Tye,
``Stability of D1-strings inside a D3-brane,''
JHEP {\bf 0403}, 055 (2004)
[arXiv:hep-th/0402072].
}

\Title{\vbox{\baselineskip12pt\hbox{hep-th/0409284}
\hbox{BRX TH-538}\hbox{PUTP-2131}\hbox{SU ITP-4/37}}}
{\vbox{
\centerline{D-terms and D-strings in open string models}
}}
\smallskip
\centerline{Albion Lawrence$^1$ and John McGreevy$^2$}
\bigskip
\centerline{$^{1}${Martin Fisher School of Physics, Brandeis 
University,}}
\centerline{{MS 057, PO Box 549110, Waltham, MA 02454-9110}}
\medskip
\centerline{$^{2}${Dept. of Physics, Princeton University,
Princeton, NJ 08544}}
\centerline{Dept. of Physics, Stanford University,
Stanford, CA 94305}
\bigskip
\bigskip

We study the Fayet-Iliopoulos (FI) D-terms 
on
D-branes in type II Calabi-Yau backgrounds.   We provide a simple 
worldsheet proof of the fact that,
at tree level, these terms only couple to scalars in closed string hypermultiplets.
At the one-loop level, the D-terms
get corrections only if the gauge group has an anomalous spectrum,
with the anomaly cancelled by a Green-Schwarz mechanism.  We
study the local type IIA model of D6-branes at $SU(3)$ angles and show that,
as in field theory, the one-loop correction suffers from a quadratic
divergence in the open string channel.  By studying the closed
string channel, we show that this divergence is related to a closed string tadpole,
and is cancelled when the tadpole is cancelled.  
Next, we study the cosmic strings that arise in the supersymmetric phases
of these systems
in light of recent work of Dvali {\it et. al.}
In the type IIA intersecting D6-brane examples, we
identify the D-term strings as D4-branes ending on the D6-branes.
Finally, we use $\CN=1$ dualities to relate these results to 
previous work on the FI D-term of heterotic strings.

\Date{September 2004}

\newsec{Introduction}

In recent years, the study of D-branes and
orientifold planes in nontrivial string backgrounds 
has been brought under a 
degree of computational control.  Within this framework, one may construct
a (bewildering) variety of supersymmetric models, some of
them with spectra close to the supersymmetric standard model. Perhaps
more interestingly, it is via D-brane models combined with magnetic fluxes
that the greatest advances have been made
towards constructing 
vacuum states
free of light scalar fields \refs{\GiddingsYU,\KachruHE,\KachruAW,
\DenefDM,\RobbinsHX,\vijayper}.

In order to address both the
practical model-building questions as well as the deeper
physical issues raised by this recent work, it is important
to have a better quantitative and qualitative understanding of
the 4d low-energy effective action (and its validity) for such models.
Most of the empirically-based puzzles of particle physics -- \eg\ the
hierarchy problem and the cosmological constant problem -- are
discussed in this arena.

In open string models, the F-terms
can be brought under control due to spacetime nonrenormalization
theorems, coupled with the fact that such terms can be computed
using topological string theory.  
The D-term couplings are notoriously more difficult to compute.
However, in these models the Fayet-Iliopoulos (FI) 
D-terms can be understood
at tree level \refs{\eg\ \DouglasGI, \AspinwallDZ}, and they
are particularly important in understanding the vacuum structure
of the theory.
Furthermore, the D-term part of the potential is central to
recent
attempts to construct inflationary models in string theory 
\refs{\DvaliPA,\AlexanderKS,\BurgessFX,\DvaliFW,
\SarangiYT,\JonesDA,\KachruSX,\CopelandBJ}.
If the inflationary
potential is dominated by D-terms, one may avoid some of the
naturalness issues that plague F-term-driven inflation
\refs{\BinetruyXJ, \HalyoPP}. 

%

There are several related puzzles regarding these terms in type II D-brane
models.  First, a field-theory nonrenormalization theorem \refs{\wf}\ states that such D-terms
get quantum corrections only at one loop order in perturbation theory,
and then only if the chiral multiplets have charges $Q_i$ under this
 $U(1)$ such that $\sum_i Q_i \neq 0$, \ie\ if the spectrum is anomalous.
 In this case, it is known that the FI D-term gets a quadratically divergent
 one-loop correction proportional to $(\sum_i Q_i) \Lambda_{UV}^2$,
 where $\Lambda_{UV}$ is the ultraviolet cutoff.  
There are string theory constructions with 
such a spectrum \refs{\DineXK, \bdl}, wherein 
the anomaly is cancelled by the Green-Schwarz mechanism 
\refs{\GreenSG, \GreenDD}.
 Our first question is, then, what is the 
stringy version of the one-loop calculation above?  In
 the heterotic string, when the anomaly is cancelled by a coupling
 $F_{\mu\nu} B^{\mu\nu}$ of the $U(1)$ gauge field strength
 to the NS 2-form potential $B$, the UV divergence is
 cut off at the string scale by the restriction of the modular integral to
 a fundamental domain of the modular group \refs{\DineXK,\ads}.
 In open string theories, as pointed out in \refs{\pop}, there is no such mechanism
 for cutting off the integral over open string moduli.  The corresponding 
cylinder amplitude is thus
 divergent.  In the type I models discussed in \refs{\pop, \bain, \ibanez}, 
the divergence is cancelled by
 a diagram with a crosscap, which is intrinsically stringy.  
One can ask how general this story is.\foot{
This question was asked by E. Poppitz and S. Kachru in 1999.}

There exist string models \bdl\ for which there is a
 single charged chiral multiplet \FayetYH, 
with an FI D-term which one can tune by hand,
 passing between phases of broken and unbroken supersymmetry \refs{\KachruVJ}.
 This is a generic local 
model \JoyceTZ\ for such transitions
in type II theories (general type IIB examples will be mirror to this).  
During such transitions, the underlying closed string background is nonsingular.
In such models
there is a BPS cosmic string whose tension is proportional to the FI D-term, 
as studied recently by 
Dvali and collaborators \refs{\DvaliZH,\BinetruyHH}.
An interesting question then arises as to the identity of this cosmic string in string theory.
As predicted by the effective field theory, there should be a tensionless string at the
phase boundary between broken and unbroken supersymmetry. 
Yet this boundary has real codimension
one and so should not correspond to some brane wrapping 
a vanishing cycle in the underlying manifold: rather, we expect 
the D-branes themselves to become singular at this 
transition \refs{\KachruVJ,\DouglasGI}.
Therefore, we should seek 
a D-brane configuration which is 
a string in spacetime, whose tension vanishes 
as the collection of 
space-filling D-branes becomes singular.

 In this paper we will answer these questions.
The outline 
is as follows.  In \S2, we
show via a worldsheet argument 
(similar to that in \BrunnerJQ) that at tree level,
the FI D-terms do not depend on
the closed-string modes which descend from $N=2$ vector multiplets.
In \S3 we study the one-loop contribution to the
D-terms for the example of two D6-branes intersecting
in $3+1$ dimensions \refs{\bdl}, and argue that
the divergent contribution cancels when all disc contributions to
massless closed string tadpoles vanishes.  In \S4, we study the
type IIA realization of the Fayet model with intersecting
D6-branes \refs{\KachruVJ,\JoyceTZ}, and study the cosmic
strings which arise in the supersymmetric phase, 
corroborating recent work
of Dvali \etal\ and Binetruy \etal\ \refs{\DvaliZH,\BinetruyHH}\
in this more nontrivial class of examples.  Our identification is
similar to a desciption given in
\refs{\HashimotoPU,\MatsudaQT,\MatsudaBK}\foot{
Recent papers which 
also study D-term strings in string theory include 
\refs{\DasguptaDW, \HalyoUU, \HalyoZJ, \LeblondUC}.}
In \S5 we conclude by discussing the relation
of our results to dual heterotic and M-theory models. 
An appendix gives the details of the one-loop open-string calculations.

In this work, we are interested in theories for which nontrivial gauge dynamics arises
from D-branes.  As in \refs{\LMone,\LMtwo}, we will focus on ``local models'', by which
we mean that we focus on low-energy D-brane dynamics 
near some region of interest of the geometry.
The low-energy dynamics can be captured by placing D-branes in a noncompact
background with the same local structure.  There are two advantages
to this approach.  First, the noncompact models are easier to describe
and decouple gravity from the problem.  Secondly,
they may be glued in to a wide variety of compact models.  Other sectors
with light fields will often be physically separated in the extra dimensions
from the sector of interest.  From the standpoint of the local model, this
will fix the behavior at infinity.

\newsec{Closed string decoupling theorems for the D-brane effective action}

For type II models at tree level, the fact that 
the closed strings lie in $\CN=2$ supermultiplets controls
how they couple to the FI D-terms: in particular
it can be shown that specific closed string modes decouple
from specific terms in the $4d$ effective action for open strings.
In the case of supersymmetric D-branes in type II string models, it
was argued in \refs{\BrunnerJQ}\ by surveying known examples
\refs{\DouglasSW,\DouglasDE}\ that FI
D-terms for open string degrees of freedom are
{\it independent} of closed string modes
which descend from $\CN=2$ vector multiplets, while superpotential
terms for open string chiral scalar multiplets are
independent of closed string modes which descend from $\CN=2$ hypermultiplets.
These facts are crucial in the
categorical description of supersymmetric D-brane configurations
\refs{\DouglasGI,\AspinwallPU}.

The statement regarding the superpotential was proven via
worldsheet techniques in \refs{\BrunnerJQ}.  We can provide
an equally simple proof of the decoupling statement for
FI D-terms at tree level.  The central point
is that $\CN=1$ spacetime supersymmetry requires
$\CN=2$ worldsheet superconformal symmetry in the open
string sector.  The $\CN=2$ algebra contains a
$U(1)_R$ affine Lie subalgebra, and the decoupling theorem
for FI D-terms is a consequence of $U(1)_R$ selection rules.
It is worth noting that in this context, 
the existence of the $U(1)$ R-symmetry
is weaker than the requirement of  unbroken 4d $\CN=1$ supersymmetry.

We are interested in the behavior of the FI D-term
under a small deformation $\delta \phi$ of the closed string background.  Let
this deformation be described by the $(-1,-1)$ picture
vertex operator $V_{\delta\phi} = e^{-\phi - \tphi} \CO$, where
$\phi,\tphi$ are the left- and right-moving bosonized superconformal ghosts
(\qv\ \refs{\joebook})
and $\CO$ is a dimension $(1/2,1/2)$ operator.
We would also like a vertex operator describing the
auxiliary field $D$ for a $U(1)$ gauge group.
Let the vertex operator for an open
string gauge field have a Chan-Paton matrix
$s_{aa'}$, where $a,a'$ label the constituent D-branes.
One may easily adapt the treatment of the heterotic string 
in \refs{\ads, \DineGJ}\ to the open string case, 
to show that the $(0)$-picture boundary operator
\eqn\auxvert{
   V_{D,aa'} = J_{U(1)} s_{aa'}
}
is the operator for the FI D-term for $U(1)_a$, 
and $s_{aa'} = \delta_{aa'}$.  In the case of the
heterotic string, $s$ is replaced by the left-moving current
for the corresponding element of the worldsheet affine Lie algebra associated
to the spacetime gauge group \refs{\DineGJ}. The variation of the
spacetime coupling $\int d^4 x \xi D$ with respect to the deformation
$\delta\phi$ is simply the disc amplitude:
\eqn\discamp{
   \langle V_D(w_0) V_{\delta\phi}(z,\bar{z})\rangle
}
where $(z,\bar{z})$ is a fixed point in the interior of the disc $D$
and $w_0$ is a location on the boundary $\p D$ of the disc.\foot{
$V_D$ is not a vertex operator for a physical state, and the
reader may feel more comfortable measuring instead
the coupling of $\delta\phi$ to the masses of scalar
fields charged under the corresponding $U(1)$, as
in \refs{\DouglasSW,\pop}.  However, 
an analogous analysis to that of 
\refs{\ads,\DineGJ} 
shows that 
the calculation described here is equivalent.}

In the models of interest, the closed string sector
has $\CN=(2,2)$ worldsheet supersymmetry, and the
spacetime fields corresponding to massless vertex
operators lie in 4d $\CN=2$ spacetime supermultiplets. 
If $\delta\phi$ denotes a scalar in a vector multiplet, 
the corresponding NS-NS vertex operator $\CO$ lies in one of the
four chiral rings.  In type IIB string theory,
these are complex structure deformations 
with $U(1)_R$ charge $(1,1)$ or $(-1,-1)$,
while in type IIA string theory they are K\"ahler
deformations with  $U(1)_R$ charge
$(\pm 1,\mp 1)$.  We are interested in D-branes
which fill the 4d spacetime and preserve supersymmetry.
As described in \refs{\BrunnerJQ}, the corresponding
open string boundary conditions preserve the diagonal combination
$J + \tilde{J}$ of left- and right-moving $U(1)_R$ current algebra
for type IIB, and the off-diagonal combination $J - \tilde{J}$
for type IIA backgrounds.  Therefore $V_{\delta\phi}$ is charged
under the preserved $U(1)_R$ current, while $V_D $ is that R-current
itself, and therefore neutral.  Therefore 
Eq.\ \discamp\ vanishes by $U(1)_R$-charge conservation.

Although it is a digression from the theme of this work, we note that
the gauge coupling for open string fields also decouples from
closed string vectormultiplets at tree level, as the proof  is identical to that for the FI D-terms.
The variation of the gauge coupling
with respect to $\delta\phi$ is proportional to the disc amplitude
\eqn\disccoupling{
   \langle V_{\delta\phi}(z, \bar{z}) V_A(w_0) \oint_{\p D} dw V_A(w) \rangle
}
where $V_A$ is the $(0,0)$-picture vertex operator for the
gauge field, and $w_0,w$ lie on the boundary of the disc.
Again, $V_A$ is neutral with respect to the preserved $U(1)_R$
symmetry of the theory, so this amplitude vanishes by
worldsheet R-charge conservation.

\newsec{One-loop FI contributions for intersecting D-branes}

Consider a general $d=4, N=1$ gauge theory with group
$G = U(1) \times G'$.  If the chiral multiplets $\Phi^i$ have
charges $q^i$, then there is a quadratically-divergent
one-loop contribution to the the FI term 
of the form \refs{\wf}:
\eqn\ficont{
   \xi_{one-loop} \propto (\sum_i q^i) \Lambda_{UV}^2
}
where $\Lambda_{UV}$ is an ultraviolet cutoff.
In other words, the FI term has a divergent contribution
when the $U(1)$ gauge symmetry is anomalous.

The standard lore is that in consistent string backgrounds, divergences are 
either cut off or have an infrared interpretation, and gauge anomalies
are cancelled via the Green-Schwarz mechanism.  The status of
the one-loop contribution to the FI D-terms seems to depend on
the model at hand.  In the case of the heterotic string,
when the anomaly is cancelled by a coupling to the
``universal'' axion dual to the NS-NS 2-form, the one-loop
contribution is finite \refs{\ads,\DineXK}, with $\Lambda_{UV}^2 = M_s^2$:
modular invariance removes the UV-divergent region
of the string diagrams. 

In the case of open string $U(1)$ gauge groups, there is no modular group to cut
off the UV region of the open string loop amplitudes.  The divergence must be cancelled or explained.
For a particular type I orentifold vacuum \refs{\pop},  the one-loop correction
was shown to vanish identically due to a cancellation between the
cylinder and M\"obius strip diagrams.  It is not clear how general
this story might be.

In order to answer this more directly, we examine
the one-loop correction to the FI D-term in a noncompact
type IIA example of two D6-branes $A,B$ in $\IR^{10}$ intersecting along
$\IR^4$, with angles chosen such that $\CN=1, d=4$ SUSY is preserved
at tree level \refs{\bdl}.  The strings localized
at the intersection of these branes lead to an anomalous spectrum
for the off-diagonal $U(1)_-$ generated by the CP matrix 
$\left(\matrix{ 1 & 0 \cr 0 & -1} \right) $.  This is a local model
of the realization in \refs{\KachruVJ}\ of the Fayet model
and provides a fairly generic picture of how FI D-terms 
for open strings arise in type IIA compactifications with branes.
In this section we compute the one-loop contribution to
the FI D-term for $U(1)_-$, and show that it indeed has
the same divergence as the field theory.  We argue that this
divergence has an infrared interpretation related to
the closed string tadpoles generated by the D-branes.  Therefore
this divergence will be cancelled when the tadpoles are cancelled,
which one must do in a consistent compact model.  


In \S5, we will use $\CN=1$ string dualities to discuss 
the relation of this result to
the heterotic string results of \refs{\DineXK,\ads}.

\subsec{Description of the model}

We will study two D6-branes in type IIA string theory, intersecting
along $\IR^4$. We write the $10d$ space as $\IR^4\times \IC^3$:
here $\IC^3$ can be thought of as a local model for a nonsingular
region of a Calabi-Yau manifold $M$.  The D6-branes will fill out three-dimensional
submanifolds of $M$.  In a general Calabi-Yau threefold, $\CN=1$ supersymmetry
requires that these submanifolds be special Lagrangian  \refs{\bbs}.  The conditions for
$\Sigma\subset M$ to be special Lagrangian can be written in terms of the
holomorphic $(3,0)$ form as:
\eqn\SLAGcond{
\eqalign{
	\Re e^{i\theta} \Omega|_{\Sigma} &= {\rm vol}~ \Sigma\cr
	\Im e^{i\theta}\Omega|_{\Sigma} & = 0
	}}
where $\theta$ is some angle, often called the ``phase'' of $\Sigma$.
In the present case, we can write a K\"ahler form
on $\CI^3$ as:
\eqn\kahlerform{
   \omega = \eta_{i\jb} dz^i d\bz^{\jb}
 }
and a holomorphic $(3,0)$ form as:
\eqn\threezero{
	\Omega = \epsilon_{ijk}dz^i dz^j dz^k
	}
Here $z^{i=1,2,3}$ are the canonical flat holomorphic coordinates on $\IC^3$.
It is easy to see that the cycles
$\Im e^{i\theta^i} z^i = 0$ are special Lagrangian cycles of $\IC^3$
with phase $\theta = \sum_i \theta_i$.  Two D6-branes wrapped on two intersecting
cycles $\Sigma_1,\Sigma_2$ with the same 
phase will together preserve $d=4,\CN=1$ supersymmetry as well, as the
union $\Sigma_1 \bigcup \Sigma_2$ satisfies the above conditions.

In our local model, we will take $\Sigma_1 \in \IC^3$ to be the submanifold
$\Im z^i = 0$, and $\Sigma_2$ to be the submanifold $\Im e^{i\theta_i} z^i = 0$.
Of course $\IC^3$ has {\it vanishing} holonomy, so the theory
has 32 supercharges instead of eight before the D-branes
are added.  We will label these branes ``1'' and ``2''  (see
fig. 1).  Still, while $\Sigma_1$ by itself preserves $\CN =4$ SUSY in four dimensions,
if we choose $\sum_i \theta_i = 0$, $\theta_i \neq 0\ \forall i$, then the branes $\Sigma_1,\Sigma_2$
together preserve $\CN=1$ spacetime SUSY in $\IR^4$ \refs{\bdl}.
The light spectrum of $4d$ fields was worked out in \refs{\bdl}.  There is a 
$U(1)^2 = U(1)_1\times U(1)_2$ gauge symmetry, and a single chiral multiplet
with charge $(1,-1)$.  The off-diagonal combination $U(1)_-$ of
$U(1)_1$ and $U(1)_2$ is therefore anomalous.  The anomaly
is cancelled by anomaly inflow due to couplings of the D-brane
to the RR potentials.
The coupling $\int_{M_6} F \wedge C^{(5)}_{RR}$ for D6-branes on $M_6$ leads
in particular to the coupling
\eqn\anomcoupling{
   \left(\int_{\IR^4\times\Sigma_1} + \int_{\IR^4\times\Sigma_2}\right)
   	F \wedge C^{(5)}
}
If $F_1,F_2$ are the gauge fields for branes $1$ and $2$, and we define
the following two-forms in $\IR^4$:
$$C_{1,2} = \int_{\Sigma_{1,2}} C^{(5)}\ ,$$
then \anomcoupling\ leads to the couplings in the 4d effective action
\eqn\fourdanom{
	S_{anom} = \half \int d^4 x \left[(F_1 - F_2) \wedge (C_1 - C_2) 
		+ (F_1 + F_2)\wedge(C_1 + C_2)\right]\ .
}
The first term on the right hand side leads to a 4d description of anomaly
cancellation via the Green-Schwarz mechanism.  It will become important
in \S4.
\ifig\angles{Two 
D6-branes intersecting along $\IR^4 \subset \IR^4 \times \IC^3$, with one
of the coordinates $z^{i=1,2,3}$ in $\IC^3$ shown.}
{\epsfxsize2.5in\epsfbox{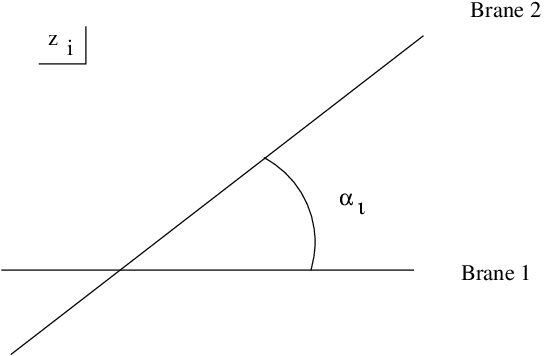}}

\subsec{Computing the one-loop contribution to the D-term.}

The diagonal linear combination $U(1)_+$ of $U(1)_1\times U(1)_2$
should get no one-loop contribution to the corresponding FI D-term,
as there are 
no charged fields are coupled to it perturbatively.
On the
other hand, 
according to the nonrenormalization theorem in \refs{\wf},
the FI D-term for the off-diagonal combination $U(1)_-$
{\it will} get a one-loop contribution in the field theory limit.  
We will now test this via
a direct computation in string theory.

In the RNS formalism, the FI D-term is proportional to the one-point function
\eqn\FIcyl{
	\delta \xi = \sum_{a,b,i} \int dt dw\ \tr_{ab,i} \half \left(1 + (-1)^F\right)
	V_{D,aa}(w) e^{-2\pi t L_0} (-1)^{F_{st}} \equiv \sum_{ab} A_{ab}
}
on the cylinder.  
The sum is over the Chan-Paton factors $a,b$ of the two boundaries, and over
the periodicity $i = ({\rm NS, R})$ of the fermions.  
$F_{st}$ is the spacetime fermion number -- this leads to a factor
of $-1$ in the Ramond sector -- and $F$ is the worldsheet fermion number.
$a$ denotes the Chan-Paton index of
the boundary at $\sigma = 0$, and $b$ the Chan-Paton index 
of the boundary at $\sigma = \pi$.  The trace is over the oscillator modes and
zero modes of the worldsheet fields.  $L_0$ the zero mode of the worldsheet
energy-momentum tensor, $t$ is the modular parameter of the cylinder, $w$ a location
on the boundary of the cylinder, and
$V_{D,ab}$ the vertex operator \auxvert\ for the FI D-term corresponding to $U(1)_-$.
For this calculation, one may either integrate
$V_D$ around the boundary of the cylinder, or divide by the
length of this boundary and fix the position of $V_D$.
Either way, one may write \FIcyl\ as
\eqn\FIcyltwo{ \delta \xi = \tr\ \int dt \half(1 + (-1)^F) (-1)^{F_{st}}V_{D,aa,0} e^{-2\pi t L_0}\ .
}
The vertex operator for the auxiliary field in the vector multiplet for $U(1)_{D}$ is:
\eqn\fivertzerodiag{
	V_{D,aa',0} = J_0 \left(\matrix{1&0 \cr 0 &1} \right)_{aa'}\ ,
}
while the vertex operator for the auxiliary field for $U(1)_-$ is:
\eqn\fivertzero{
	V_{D,aa',0} = J_0 \left(\matrix{1&0 \cr 0 &-1} \right)_{aa'}\ .
}
Here $J_0$ is the zero mode of the $U(1)_R$ current,
and the matrix acts on the Chan-Paton indices $a=1,2$ of the boundary at $\sigma=0$
in which $V_D$ is inserted.  The indices $a=1,2$ denote branes $1$ and $2$,
respectively. (See fig. 2.)
\ifig\cylinder{The one-loop diagram for $V_D$.  Here $a,b=1,2$ labels
whether each boundary ends on brane $1$ or $2$, respectively.}
{\epsfxsize3.5in\epsfbox{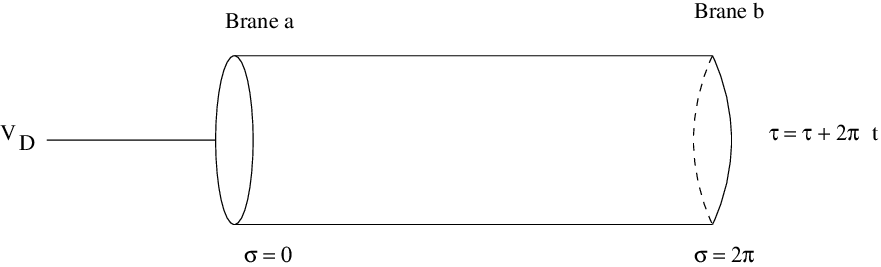}}
We will find it useful to rewrite the amplitude \FIcyl\ as 
\eqn\FIderivative{
\eqalign{
\vev{ V_D}  & =
{-i \over 2\pi} \partial_{\nu}\ \  \sum_{abi} \tr_{ab,i} \int dt  \half(1 + (-1)^F) (-1)^{F_{st}}
	s_{aa}
	e^{2\pi i \nu J_0} e^{-2\pi t L_0}|_{\nu=0}\cr
	&  \equiv {- i\over 2\pi} \p_{\nu}\ \ 
	\tr \int dt\ \half(1+(-1)^F) (-1)^{F_{st}} s_{aa}z^{J_0} q^{L_0}|_{\nu=0}\ ,
}}
where
$$ z \equiv e^{2\pi i \nu}  \ ;\ \ \ \ q\equiv e^{-2\pi t} \equiv e^{2\pi i \tau}\ . $$

We can break up the amplitude into Chan-Paton sectors:
\eqn\sectorsum{
	\eqalign{
	\delta\xi_{+} & = {-i \over 2\pi} \p_{\nu}\ \int dt \left(A_{11}(t) + A_{12}(t) 
		+ A_{21}(t) + A_{22}(t)\right) \cr
	\delta\xi_{-} & = {-i \over 2\pi} \p_{\nu}\ \int dt \left(A_{11}(t) + A_{12}(t) 
	- A_{21}(t) - A_{22}(t)\right)\ ,
	}}
where $\xi_{+}, \xi_{-}$ are the FI D-terms for $U(1)_+$ and $U(1)_-$, respectively.
We will now discuss each of the $A_{ab}$s in turn.

\vskip .3cm
\noindent $A_{11}$ {\it and} $A_{22}$.

Strings in the $11$ and $22$ sectors are neutral with respect to both $U(1)_1$ and
$U(1)_2$.  Therefore, we expect them to give a vanishing contribution to the
FI D-terms.  We can see this by simply examining the trace over fermion
modes in the $\IC^3$ direction.  The point is that these fermions are the
only fields on the worldsheet charged under the worldsheet
$U(1)_R$, and so lead to all of the $\nu$-dependence of the amplitudes. 
For the $11$ sector, we can use the mode
expansions described in Appendix A to find:
\eqn\fermiontraces{
\eqalign{
	\tr_{11,NS} (-1)^{F_{st}} z^{J_0} q^{L_0} & = 
	\prod_{n=1}^{\infty} (1 + z q^{n-\half})^3 (1 + z^{-1} q^{n-\half})^3\cr
	\tr_{11,NS} (-1)^{F+F_{st}} z^{J_0} q^{L_0} &=
	\prod_{n=1}^{\infty} (1 - z q^{n-\half})^3 (1 - z^{-1} q^{n-\half})^3\cr
	\tr_{11,R} (-1)^{F_{st}}z^{J_0} q^{L_0} &=
	- (z^{1/2} + z^{-1/2}) \prod_{n=1}^{\infty} (1 + z q^{n})^3 (1 + z^{-1} q^{n})^3\ .
	}}
The expressions for the $22$ sector are identical.
We do not write out $\tr_{11,R}(-1)^{F+F_{st}} (\ldots)$ because the spacetime
fermion zero modes lead to a vanishing contribution from this sector.

All of the expressions in \fermiontraces\ are even under $\nu\to -\nu$,
and regular at $\nu = 0$.  Therefore, the derivative of any of these expressions
with respect to $\nu$ must vanish at $\nu = 0$, and so $A_{11} = A_{22} = 0$.
	
\vskip .3cm
\noindent $A_{12}$ {\it and} $A_{21}$.

The strings stretching between branes $1$ and $2$ are charged under $U(1)_-$.
We expect any contributions to $\delta\xi_{-}$ to come from the $12$ and $12$ 
Chan-Paton sectors.
We will work through the computation of $A_{12}$ in some detail.  It will be clear from
the form of the answer that $\p_{\nu}A_{21} = - \p_{\nu}A_{12}$.

Let us begin with the trace over the spacetime fields and ghosts.  The trace over the
superconformal ghosts will cancel the trace over the fermion modes and 
bosonic oscillator modes for the ``longitudinal'' fields $X^0,X^1,\psi^0,\psi^1$ (see Appendix A
for notation).  The contribution to $A_{12}$ from the spacetime bosons and the
bosonic ghosts is:\foot{We use the 
conventions in \joebook\ for $\eta(q)$ and for the theta functions $\vartheta_{ab}$.}
\eqn\stbosonictrace{
	A_{12,{\rm s.t.} {\rm bosons}} =
	\int {d^4 p\over (2\pi)^4} e^{-\pi t p^2/2}
	{q^{{1\over 12}} \over \eta(\tau)^2} q^{E_{gs,1}} = 
	\left({1\over \pi^2 t}\right)^2 {q^{{1\over 12}} \over \eta(\tau)^2}\ ,
	}
where $E_{gs,1}$ is the vacuum energy of these fields.
The nonzero contributions of the spacetime fermions and fermionic ghosts are:
\eqn\stfermiontrace{
\eqalign{
	 \tr_{12,NS,{\rm s.t.} {\rm fermions}} z^{J_0} q^{L_0} 
	&= q^{{1\over 24} + E_{gs,3,NS}} {\vartheta_{00}(0,\tau) \over \eta(\tau)} \cr
	\tr_{12,NS,{\rm s.t.} {\rm fermions}}(-1)^{F} z^{J_0}q^{L_0} &=
	q^{{1\over 24} + E_{gs,3,NS}} 
	{\vartheta_{01}(0,\tau)\over \eta(\tau)} \gamma_{3} \cr
	\tr_{12,R, {\rm s.t.} {\rm fermions}} z^{J_0} q^{L_0} & =
	- q^{-{1\over 12} + E_{gs,3,R}} {\vartheta_{10}(0,\tau)\over\eta(\tau)}\ ,
}}
where $E_{gs,3,NS}$ and $E_{gs,3,R}$ are the ground state energies of these
modes in the Neveu-Schwarz and Ramond sectors, respectively, and
$(-1)^F |0,NS\rangle = \gamma_{3} |0,NS\rangle$ in the NS sector.

The contribution of the ``internal'' complex
worldsheet bosonic fields $Z^{i=1,2,3}$ is:
\eqn\intbostrace{
	A_{12,{\rm int.} {\rm bosons}} = 
		(-i)^3 q^{1\over 4} {\eta(\tau)^3 \over \prod_{i=1}^3
		\vartheta_{11}(|\alpha_i| \tau,\tau)} q^{E_{gs,2}}\ ,
	}
where $E_{gs,2}$ is the vacuum energy of these fields.  As in Appendix A,
$\alpha_i = {\theta_i \over 2\pi}$, where $\theta_i$ is the angle between
branes $1$ and $2$ in the $z^i$-plane of $\IC^3$.

Finally, the contribution of the ``internal'' complex fermions is:
\eqn\intfermiontrace{
\eqalign{
	\tr_{12,NS,{\rm int.} {\rm fermions}}
	(-1)^{F_{st}}z^{J_0} q^{L_0} & = 
	q^{{1\over 8} + E_{gs,4,NS}}
	{\prod_{i=1}^3 \vartheta_{00}(\nu + \alpha_i\tau,\tau)\over \eta(\tau)^3}\cr
	\tr_{12,NS,{\rm int.} {\rm fermions}}
	(-1)^{F+F_{st}}z^{J_0} q^{L_0} & = 
	q^{{1\over 8} + E_{gs,4,NS}}
	{\prod_{i=1}^3 \vartheta_{01}(\nu + \alpha_i\tau,\tau)\over \eta(\tau)^3}
	\gamma_4\cr
	\tr_{12,R,{\rm int.} {\rm fermions}}
	(-1)^{F_{st}}z^{J_0} q^{L_0} & = 
	q^{-{1\over 4} + E_{gs,4,R}}
	{\prod_{i=1}^3 \vartheta_{10}(\nu + \alpha_i\tau,\tau)\over \eta(\tau)^3}
	}}
Again, $E_{gs,4,NS}$ and $E_{gs,4,R}$ are the ground state energies
in the NS and R sectors, respectively, and 
$(-1)^F |0,NS\rangle = \gamma_4 |0,NS\rangle$.

Putting these all together, we find that:
\eqn\totalamplitude{
\eqalign{
	A_{12} & = \left( {1\over \pi^2 t}\right)^2 {(-i)^3\over 
	\eta(\tau)^3 \prod_{i = 1}^3 \vartheta_{11}(\alpha_i\tau,\tau)}
	\left[ \vartheta_{00}(0,\tau)\prod_{i=1}^3\vartheta_{00}(\alpha_i\tau,\tau)\right.\cr
	& \left.- \vartheta_{01}(0,\tau)\prod_{i=1}^3\vartheta_{00}(\alpha_i\tau,\tau)
	- \vartheta_{10}(0,\tau)\prod_{i=1}^3\vartheta_{00}(|\alpha_i|\tau,\tau)
	\right]
}}
Here we have used the fact that $(-1)^F|0,NS\rangle = 
\gamma_3\gamma_4 \ket{0,NS} =- |0,NS\rangle$,
that $\sum_{k = 1}^4 E_{gs,k,NS} = -\half$, and that 
$\sum_{k=1}^4 E_{gs,k,R} = 0$.

This expression can be simplified via the Riemann theta identities
(\qv\ Eq. (13.4.20-21) of \refs{\joebook}\ or chapter 1,\S5, of \refs{\tatatheta})
to:
\eqn\totalamplitudesimp{
	A_{12} = \left( {1\over \pi^2 t}\right)^2 { (-i)^3
	\vartheta_{11}({3\nu\over 2},\tau) \prod_{i=1}^3
	\vartheta_{11}(-{\nu\over 2} + \alpha_i\tau,\tau)\over 
	\eta(\tau)^3 \prod_{i = 1}^3 \vartheta_{11}(|\alpha_i|\tau,\tau)}\ .
	}
Next, we wish to compute $\p_{\nu} A_{12}|_{\nu=0}$.  Because
$\vartheta_{11}(0,\tau) = 0$, 
\eqn\totalamplitudesimp{
	{- i\over 2\pi} A_{12}|_{\nu=0} = \left( {1\over \pi^2 t}\right)^2 {
	(\p_{\nu} \vartheta_{11}({3\nu\over 2},\tau))_{\nu=0} \prod_{i=1}^3
	\vartheta_{11}(\alpha_i\tau,\tau)\over 
	2\pi \eta(\tau)^3 \prod_{i = 1}^3 \vartheta_{11}(\alpha_i\tau,\tau)}
	= {18\over\pi t^2}\ .
	}
where we have used $\partial_\nu \vartheta_{11}(\nu,\tau)|_{\nu=0} = (-2\pi\eta(\tau))^3$.
We have also used the fact that 
one of the angles $\alpha$ is negative, which we 
have chosen to be $\alpha_3 < 0$: in this case,
$$ {\vartheta_{11}(\alpha_3\tau,\tau) \over \vartheta_{11}(|\alpha_3|\tau,\tau)} = -1\ , $$
which contributes an additional minus sign, leading to the overall sign in \totalamplitudesimp.

Inspection of \intfermiontrace\ reveals that 
$A_{12}$ is invariant under the combined operation
$\alpha_i\to-\alpha_i$, $\nu\to-\nu$.   In the mode expansion
for the $21$ sector, the only difference from the $12$ sector
is that the angles have the opposite sign.  One may therefore write
$A_{21}(\nu,t) = A_{12}(-\nu,t)$, and so
\eqn\flipsector{
	\partial_{\nu} A_{21}|_{\nu=0} = -\partial_{\nu} A_{12}|_{\nu=0}\ .
	}

\subsec{Physical interpretation of the one-loop amplitudes for $\delta\xi_{+,-}$}

Using Eq. \totalamplitudesimp\ and \flipsector, we find:
\eqn\FIanswer{
\eqalign{
	\delta \xi_{+} &= {-i \over 2\pi} \partial_{\nu} \int dt\
		\left(A_{12} + A_{21}\right) = 0 \cr
	\delta \xi_{-} &= {-i \over 2\pi} \partial_{\nu} \int dt\
		\left(A_{12} - A_{21}\right) = {36\over \pi} \int^{\infty}_{\epsilon} dt {1 \over t^2}\ ,
}}
where $\epsilon = \Lambda_{UV}^{-2}$ is the open-string-channel 
ultraviolet cutoff.
These answers are consistent with the results of \refs{\wf}.  
As expected, there is
no correction to the FI D-term for $U(1)_+$, since there is no
massless chiral multiplet charged under this $U(1)$. For $\delta\xi_{-}$,
the form of  \totalamplitudesimp\ indicates that none of the oscillator modes
contribute.  
The fact that the oscillator contributions have cancelled 
reflects the 
fact that massive fields do not renormalize the FI term;
this is an index quantity. 
The zero modes for the massless chiral multiplet charged
under $U(1)_-$ lead to a quadratic ultraviolet divergence.

Unlike the heterotic string, this divergence is not cut off by any modular 
group action.
Such a divergence must have an infrared interpretation.  Indeed, if one performs
a modular transformation $\tau_{cl} = - {1\over\tau} = {i\over t}$ to the
closed string channel (see fig. 3), then we find that
\eqn\closedchannel{
	\delta\xi_{-} = {- 36 i \over \pi} \int_{0}^{i \epsilon} d\tau_{cl}
}
This represents an infrared divergence in the closed string channel.  This
divergence is due to the exchange of
massless closed string modes in the factorization limit illustrated in fig.~3.  
The divergence from $A_{12}$ ($A_{21}$) is proportional to the tadpoles
generated by D6-brane~$2$ (D6-brane~$1$).  It will therefore be cancelled when
the tadpoles are cancelled\foot{
In noncompact examples, this can be accomplished by 
solving the equations of motion for the RR 7-form potential in the 
presence of the D6-brane source.  We thank Atish Dabholkar 
and Howard Schnitzer for bringing this fact to our attention.}.  
This is consistent with the result of \refs{\pop}.

\ifig\closedchannel{Factorization of the one-loop correction to $\xi_{-}$.}
{\epsfxsize3.0in\epsfbox{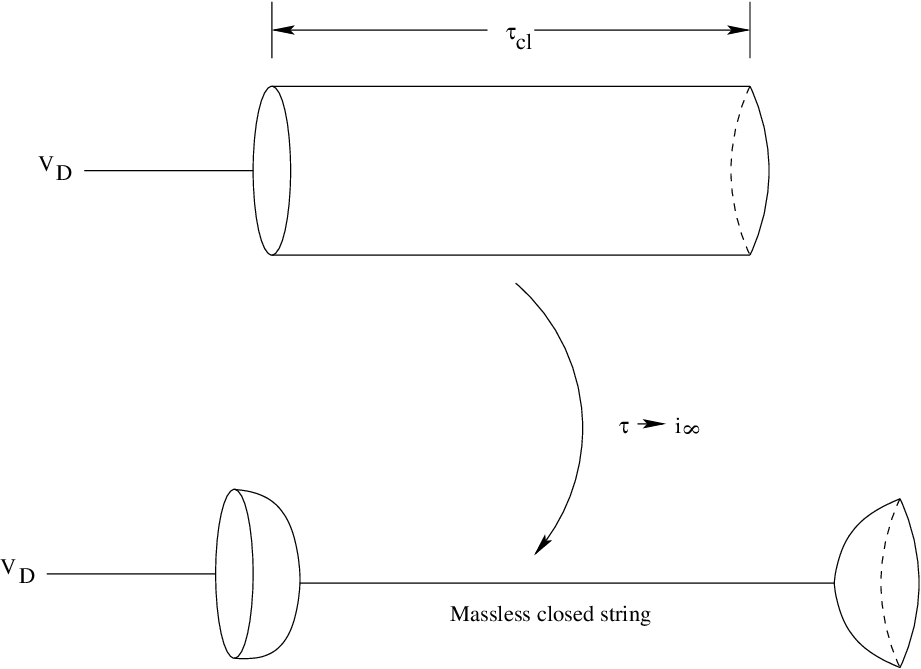}}

A remaining question is whether any finite correction to $\xi_{-}$ can remain.
If the D-brane tadpoles are cancelled by perturbative orientifolds, the
correction will vanish exactly, as in \refs{\pop}, if the orientifold projection
preserves SUSY.  The nonrenormalization theorem of \refs{\wf},
and the form of the string amplitude \totalamplitudesimp, imply that
the contribution of each massive supermultiplet vanishes separately.
This will remain true for fields which survive the orientifold projection.
The orientifold sectors will give no contributions
to $\xi_{-}$ from open string oscillator modes, and the divergence in these
sectors has the same form $c \int dt/t^2$ for some constant $c$.  
If the orientifold cancels the divergence, there is no room for a finite part.
This is a mystery from the standpoint of the low-energy effective field theory,
unless there is a symmetry which would forbid such a term.  
It is tempting to blame the 
underlying $\CN=2$ structure 
of the closed-string physics; we
leave this question for future work.

\newsec{Cosmic D-term strings are D-branes}

\lref\AspinwallJR{
P.~S.~Aspinwall,
``D-branes on Calabi-Yau manifolds,''
arXiv:hep-th/0403166.
}

The local model we studied in the previous example is one where the FI D-term
is vanishing at tree level.  The local model in type IIA
for an open string system with a single charged chiral scalar field,
and tree-level FI D-term
near zero, has been described mathematically by
\refs{\JoyceTZ}\ and given the physical interpretation we will use in
\refs{\KachruVJ} (see also \refs{\AspinwallJR} for a nice review).
As in the previous section, the low-energy gauge group
of the model is $U(1)_1\times U(1)_2 = U(1)_+ \times U(1)_-$.  There is a chiral superfield
$\Phi$ with scalar component $\phi$, 
charged under $U(1)_-$.  The D-term potential is:
\eqn\dtermpot{
	V_D = e^2 \left( |\phi|^2 - \xi\right)^2\ .
}
When $\xi < 0$, $V_D \neq 0$ and SUSY is broken; meanwhile, at the minimum of $V_D$,
$\phi = 0$ and both $U(1)_+$ and $U(1)_-$ are unbroken.  When $\xi > 0$,
there is an $S^1$ vacuum manifold at $|\phi|^2 = \xi$. SUSY is unbroken,
but $U(1)_-$ is Higgsed in the vacuum.

This setup can be realized by D6-branes in type IIA, as described by \refs{\JoyceTZ,\KachruVJ}.
The D6-branes wrap various special Lagrangian three-cycles $\Sigma \subset M_6$,
where $M_6$ is a local model of a Calabi-Yau background.
The geometric description of the system is show on the left in fig. 4.  $U(1)_1$ is
the gauge group for a D6-brane wrapped on the three-cycle  $N_+$ 
and $U(1)_2$ is the gauge group for a D6-brane on cycle $N_-$.  
$\xi$ is determined by complex structure moduli.
For $\xi < 0$, the cycles $N_{\pm}$
have different phases in the sense of \SLAGcond.  As $\xi \to 0$, the
strings stretching between $N_+$ and $N_-$ include a light chiral multiplet
charged under $U(1)_-$.  At $\xi = 0$, this multiplet becomes massless, and
SUSY is restored: the phases of $N_{\pm}$ become identical.  When
$\xi > 0$, the lightest chiral scalar becomes tachyonic.  
If one condenses this scalar field, the two three-cycles
$N_{\pm}$ merge into the three-cycle $N$ which is equivalent in homology to $N_+ + N_-$.

In the $\xi > 0$ phase, there should be cosmic string solutions, which we will call
``D-term strings''.  Dvali and collaborators
\refs{\DvaliZH,\BinetruyHH}\ have shown that such solutions have the following properties.

\item{1.}  The D-term strings have tension proportional $T = 2\pi \xi$.

\item{2.} The D-strings are BPS, with chiral $\CN=(2,0)$ worldsheet supersymmetry.

\item{3.} The D-term string should be a magnetic flux tube for
the anomalous $U(1)$ under which $\phi$ is charged.
If this anomaly is cancelled via coupling to an axion $a$:
\eqn\axioncoupling{
	A_{-}^\mu \p_{\mu} a
}
with $a \sim a + 2\pi$, then $a$ should shift by $2\pi$ around an $S^1$ which circles
the D-term string once.

Dvali \etal\ \DvaliZH\ argued that the cosmic strings which can appear in 
$D$-$\bar{D}$ brane annihilation,
such as in ``brane inflation'' scenarios 
\refs{\DvaliPA-\CopelandBJ},
are in 
fact D-term strings.  More generally, they showed that the inflaton potential energy
in such scenarios has a D-term component.  These are potentially
advantageous models.  SUSY is broken during inflation, and
if F-terms contribute significantly to the inflaton potential,  this breaking
makes the inflaton dangerously heavy unless the theory is somewhat finely tuned
\refs{\CopelandVG,\StewartTS}.  If the inflaton potential arises from an FI D-term and F-terms
do not contribute significantly, no fine-tuning is 
required in order to keep the inflaton potential flat \refs{\BinetruyXJ, 
\HalyoPP}.

In the case that D-term inflation arises from brane-antibrane
annihilation, the D-term $\xi$ scales inversely
with the string coupling itself, and vanishes only at strong coupling.  
For D6-branes in type IIA Calabi-Yau backgrounds (the only branes filling out $\IR^4$
which potentially preserve $\CN=1, d=4$ SUSY), the scenario 
described above is more general,
in the sense that the FI coupling can be a function of all of the 
hypermultiplet moduli.
The basic issue, as systematized by Douglas \refs{\DouglasGI}, 
is that for D-branes in Calabi-Yau backgrounds, the notion of
``brane'' and ``antibrane'' depends on the closed string fields descending 
from $\CN=2$ hypermultiplets -- complex structure moduli, in this case.  
Therefore, if one can stabilize these moduli along the lines of
\refs{\KachruAW}\ with some control over their vevs, one has great freedom in designing
D-term potentials
\foot{In general the F-terms introduced by stablization of the 
K\"ahler moduli
can overwhelm the D-terms \refs{\ArkaniHamedMZ,\BinetruyHH}; one must take
care in constructing actual models.}.  
As a brane inflation scenario, this is the embedding into a Calabi-Yau background
of the scenario in \refs{\GarciaBellidoKY,\JonesDA}.

A natural question is, then, the identity of the D-term strings in these 
models.  
A puzzle is that a natural way to get a light cosmic string would be by 
wrapping a $(p+1)$-brane around a vanishing
$p$-cycle $\Sigma_p \subset M_6$.  But the transitions described in 
\refs{\JoyceTZ,\KachruVJ,\DouglasGI}\ 
generically occur in perfectly regular interior points of the closed-string 
moduli space.

Inspired by \refs{\MatsudaQT,\MatsudaBK},
we identify the D-term strings  for the model described in fig. 4 in the following way.
As $\xi \to 0_+$, the SUSY 3-cycle develops a pinch; the local geometry near the
pinch is a ``Lawlor neck'' (\qv\ \refs{\JoyceTZ,\AspinwallJR}), topologically $S^2\times\IR$.
Each $S^2$ bounds a 3-ball.  There is a minimal-volume, 
special Lagrangian 3-ball,
called $D$ in fig. 4, which is bounded by the smallest $S^2$ $S = \p D$ in the neck.  
$D$ goes to zero volume as $\xi \to 0$, and it has phase $i$, in the sense of \SLAGcond,
relative to $N$.  Now, a D4-brane can consistently end on a D6-brane along a submanifold
of codimension $3$.  Therefore, a D4-brane with worldvolume 
$D\times \IR^{1,1} \subset M_6\times \IR^4$ is a candidate for the D-term string, as it
has vanishing tension in the four-dimensional theory when $\xi \to 0, {\rm vol}(D)\to 0$.
This is consistent with property (1) above.  In the remainder of this section we will
show that this string has properties (2),(3) listed above, and so is a good candidate for
the D-term string in the model of \refs{\JoyceTZ,\KachruVJ}.
\ifig\higgsphase{The local geometry of the intersection in the $\xi < 0$ broken-SUSY 
phase, the $\xi = 0$ phase boundary, and the $\xi > 0$ Higgs phase.}
{\epsfxsize2.5in\epsfbox{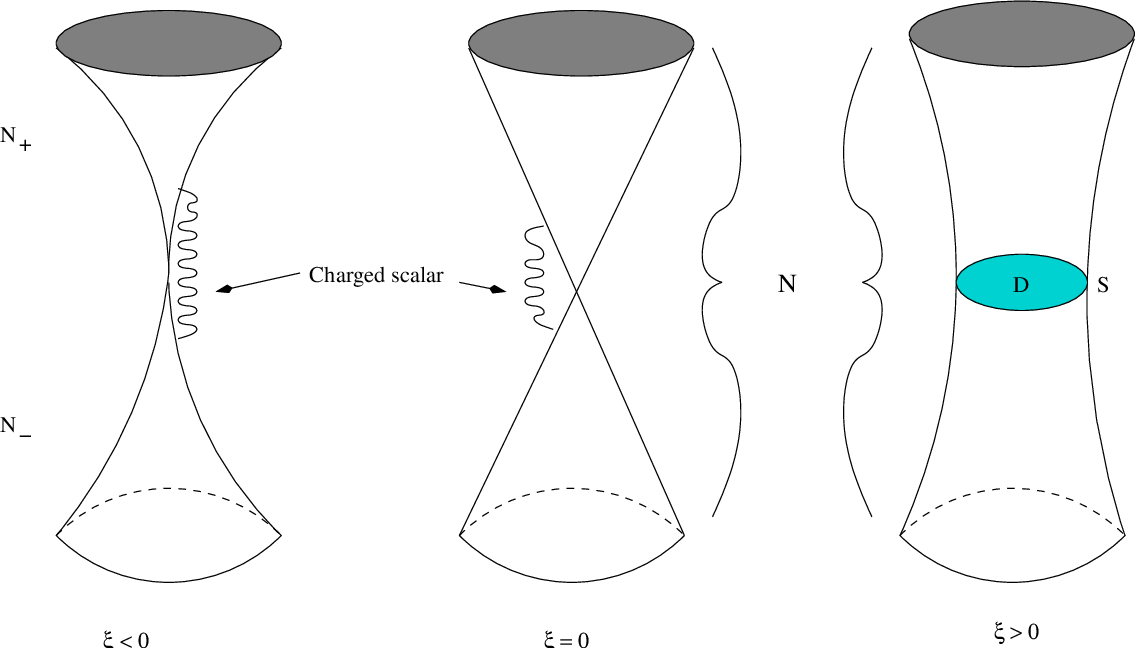}}

\subsec{Worldsheet SUSY for the D-string}
    
\lref\OoguriCK{
H.~Ooguri, Y.~Oz and Z.~Yin,
``D-branes on Calabi-Yau spaces and their mirrors,''
Nucl.\ Phys.\ B {\bf 477}, 407 (1996)
[arXiv:hep-th/9606112].
}
                 
Choose the overall phase of the holomorphic three-form
$\Omega$ so that 
$N$
has phase 0, \ie\ $\Im \Omega|_N = 0$.
Let $\Sigma$ be the spectral flow operator on for the open string
which generates $\half$-unit of spectral flow from $NS \to R$.
The operator generating a full unit of spectral flow
$NS \to NS$ is
\eqn\unitspecflow{
   \Sigma^2 = \Omega_{ijk}\psi^i\psi^j\psi^k
}
The boundary conditions on $\psi,\Omega$ then imply that
$\Sigma_+ = \bar{\Sigma}_-$ where $+,-$ denote left- and right-movers
as in Appendix A.  This boundary condition is consistent
with some general facts regarding the $\CN=2$ worldsheet CFT.
The $U(1)_R$ current can be written in terms of a worldsheet boson,
$J_{R,\pm} = c \p_{\pm} H$.  We can then write $\Sigma_{\pm} = e^{i a H_{\pm}}$
For the A-type boundary conditions required for  the D6-branes to preserve
$\CN=1$ SUSY in $d=4$ \refs{\bbs,\OoguriCK}, $J_{R,+} = - J_{R,-}$.

Next, let $S_{\alpha,\pm}$ be the spin fields for the spacetime fermions, so that
the currents implementing $\CN=2$ spacetime SUSY for the closed string theory are:
\eqn\spacetimecurrents{
\eqalign{
   Q_{\alpha,+} &= e^{-\phi/2} S_{\alpha,+} \Sigma_+ \cr
   Q_{\alpha,-} & = e^{-\tilde{\phi}/2} S_{\alpha,-} \bar{\Sigma}_{-}\ .
 }}
If all directions in spacetime are Neumann, then one can show that
$S_{\alpha,+} = S_{\alpha, -}$.  The boundary conditions on $\Sigma_{\pm}$ then
imply that 
\eqn\preservedSUSY{Q_\alpha = e^{- \phi /2} \Sigma_+ S_{\alpha, +} - e^{-\tilde{\phi}/2}
 \bar{\Sigma}_- S_{\alpha, -}}
and the corresponding operator for $\bar{Q}_{\dot{\alpha}}$
generate the $\CN=1$ spacetime supersymmetry preserved by the D6-branes.

We would now like to understand which of these SUSYs are preserved
in the presence of our candidate D-term string. 
Therefore, $\Sigma_L = -i \bar{\Sigma}_R$ for strings ending on this D4-brane.  
$Q_\alpha$ in \preservedSUSY\ is then preserved 
if $S_{\alpha,L} = i S_{\alpha, R}$.  This is in fact a consequence of
worldsheet SUSY and the boundary conditions on the spacetime coordinates
for strings ending on the D4-brane.  In spacetime, one can write the spacetime
spin fields as:
\eqn\spinfields{
\eqalign{
	S_{1,{\pm}} &= e^{i(H_{1,\pm} + H_{2,\pm})/2}\ ,\ \ \ \ 
		S_{2,\pm} = e^{-i(H_{1,\pm} + H_{2,\pm})/2}\cr
	S_{\dot{1},{\pm}} &= e^{i(H_{1,\pm} - H_{2,\pm})/2}\ ,\ \ \ \ 
		S_{\dot{2},\pm} = e^{-i(H_{1,\pm} - H_{2,\pm})/2}
}}
where
\eqn\bosonize{
\eqalign{
      e^{i H_1} &= \psi^t + i \psi^z \cr
      e^{i H_2} &= \psi^2 = \psi^x + i \psi^y
}}
Here $t,x,y,z$ are coordinates on $\IR^4$. 
Assume that the string is stretching along $z$.  $\psi^{t},\psi^{z}$
are the worldsheet superpartners of scalars with
Neumann boundary conditions, and $\psi^x,\psi^y$ are
superpartners of a scalars with
Dirichlet boundary conditions, so
\eqn\spacetimestringfermbc{
	 \psi^{t,z}_+ = \psi^{t,z}_-\ ;\ \ \ \ \ \psi^{x,y}_+ = - \psi^{x,y}_-\ .
 }  
 This boundary condition is then consistent with the boundary condition
 $e^{i H_{1,+}} = e^{i H_{1,-}}$, $e^{i H_{2,+}} = i e^{i H_{2,-}}$.  
In terms
 of the spin fields, this implies:
\eqn\bcspinfields{
\eqalign{
	S_{1,+} & = i S_{1,-}
	\ ;\ \ S_{2, +} = - i S_{2,-} \cr
	S_{\dot{1},+} & = -i S_{\dot{1},-}
	\ ;\ \ S_{\dot{2}, +} =  i S_{\dot{2},-}
}}
This means that of the four supercharges in \preservedSUSY\ and its conjugate,
$Q_{1}, Q_{\dot{2}}$ are preserved.  These supersymmetries are both right-moving
along $z$.  Therefore the D4-brane on $D\times\IR^{1,1}$ has $\CN=(2,0)$
SUSY in $\IR^{1,1}$, consistent with our identification of this
D4-brane as the D-term string.

\subsec{Magnetic flux and  axion charge carried by the D-string}

\lref\StromingerAC{
A.~Strominger,
``Open p-branes,''
Phys.\ Lett.\ B {\bf 383}, 44 (1996)
[arXiv:hep-th/9512059].
}

As stated above, for the D4-brane on $D\times\IR^{1,1}$ to be the
D-term string we claim it is, it should carry magnetic flux under the
anomalous gauge field; in other words, $\int_{\IR^2} F_{-} = 1$,
where $\IR^2$ is the plane transverse to the string in four dimensions.
When the anomaly in $U(1)_-$ is cancelled via the Green-Schwarz mechanism
through the coupling \axioncoupling, this 
is gauge-equivalent to the statement that 
the string should have axion charge.
This axion charge can be seen as follows.
The specific coupling which takes the form \axioncoupling\
is the dimensional reduction of the Wess-Zumino coupling \refs{\GreenDD}
\eqn\gscoupling{
	S_{7,GS} = \sum_i \int_{N_i\times \IR^4} F_i \wedge C^{(5)}
}
where $F_i$ is the worldvolume 
gauge field strength on the $i$th D6-brane, 
$C^{(5)}$ the 5-form RR potential of type IIA
string theory, and $\{N_i\}$ is the collection of D6-branes.
If the D6-branes wrap
$N = N_+ + N_-$, and 
for modes of $F$ which
 are independent of the internal space,
and are polarized along $\IR^4$,
\gscoupling\
can be written as\foot{
This is completed to the object
\eqn\stuckelberg{
L\ni ( \del a_- + A_- ) ^2 
}
which is invariant under the gauge transformation
\eqn\gaugetransf{
a \mapsto a - \lambda, ~~A_\mu \mapsto A_\mu - \del_\mu \lambda .}
In unitary gauge, $a_-=0$, and \stuckelberg\ is a mass term for 
the relative gauge field $A$.  
}
\eqn\fourdgs{
\eqalign{
	S_{4, GS} & = \half \int d^4 x 
	\left[(F_+ - F_-) \wedge (C^{(2)}_+ - C^{(2)}_-)+
	(F_+ + F_-) \wedge (C^{(2)}_+ + C^{(2)}_-)\right]\cr
	& = \half \int d^4 x \left( A_{-}^\mu \p_{\mu} a_{-} + 
	A_{+}^\mu \p_{\mu} a_{+} \right)
}}
Here $C^{(2)}_{\pm} = \int_{N_{1,2}} C^{(5)}$, and in four dimensions,
$da_{\pm} = \star_4 d(C^{(2)}_{+} \pm C^{(2)}_-)$.  
If the D4-brane on $D\times\IR^{1,1}$ is the D-term 
string we claim it is, the worldvolume $\IR^{1,1}$ should couple to 
$C^{(2)}_+ - C^{(2)}_-$, so that $a_{-}$ should shift by $2\pi$ around
any $S^1$ encircling the string in $\IR^4$.  Because of \fourdgs,
this will be true if the string carries magnetic flux under $U(1)_-$.
\ifig\open{The ingredients involved 
in the construction of $\Sigma_{{\rm in, out}}$, 
used to deduce the magnetic flux and axion
charge carried by the string.}
{\epsfxsize3.0in\epsfbox{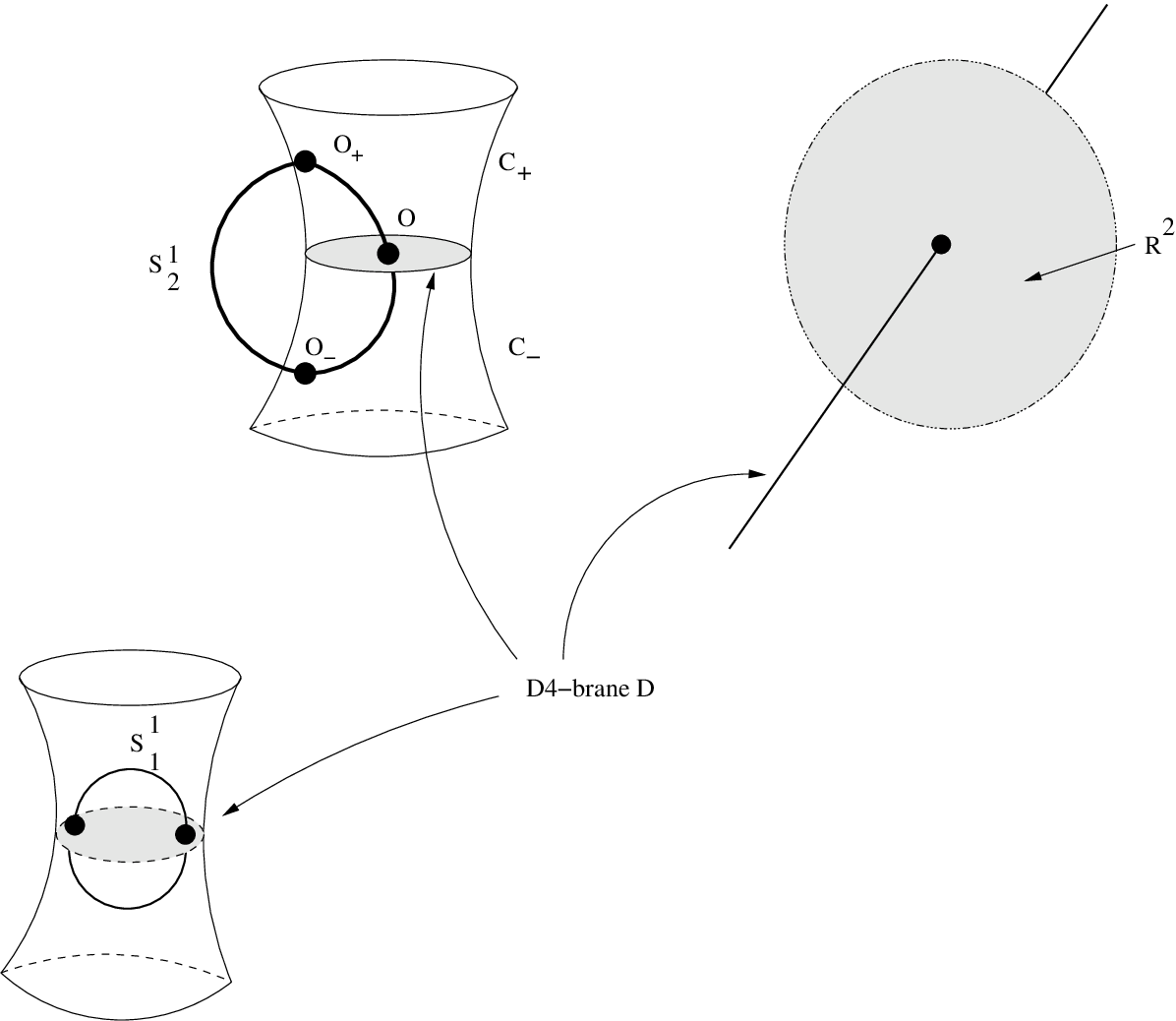}}
We can find the magnetic flux carried by our D-string by
adapting the arguments in \refs{\StromingerAC}.  The geometric setup is 
shown in \open.  
One must imagine an extra $S^2$ at each point in the figure, so that
$D$ is a three-ball, and its boundary a 2-sphere in $N$.  Finally,
there is an additional two-manifold 
$K \subset \IC^3$ which is transverse to both
$N,D$, and which is in general noncompact.  
Pick a point $P \in K$ lying on $D$.  
 
We can construct a 5-cycle $\Sigma$ by gluing the following 5-chains 
along their boundaries:
\item{1.} $\Sigma_{{\rm in}}$.  
Take the spacelike disc $B_1\subset \IR^{3,1}$, 
which intersects the
string once in spacetime at a point $Q$, 
and $K_1 \subset K$ which is a disc with $P$ in the interior.
Finally, imagine a family $\gamma_t$ of circles 
which interpolates from $S_2^1$ at
$t = 0$ to $S_1^1$ at $t = 1$.  Define $\Sigma_{{\rm in}}$ 
as an $S^1$ fibration of
$B_1 \times K_1$, such that the $S^1$ is $\gamma_0$ at 
$(Q,P)\in B_1\times K_1$,
and $\gamma_1$ at the boundary $
\del (B_1 \times K_1)
= \del B_1\times K_1 \cup B_1 \times \del K_1 $.
As indicated in \open, this
has positive intersection with 
the D4-brane worldvolume, $\Delta$ at $(P,Q,O)$, 
positive intersection with $C_+$ at
$(P,Q,O_+)$, and negative intersection with $C_-$ at $(P,Q,O_-)$.
It is called $\Sigma_{{\rm in}}$ because 
the D4-brane passes through it.

\item{2.} $\Sigma_{{\rm out}}$.  Now take the disc $B_2$ 
which intersects the string at $Q'$,
with the boundary $\p B_2 = - \p B_1$, so that $B_1$ and 
$B_2$ can be glued together to form a 2-sphere.  
Similarly, take $K_2$ to be a disc with $\p K_2 = - \p K_1$.
Let $\Sigma_{{\rm out}} = B_2 \times K_2 \times \gamma_1$.  This intersects 
$\Delta$ twice, but each time
with opposite sign, so the total intersection is zero.  

\noindent
Now, $\Sigma \equiv \Sigma_{{\rm in}} \cup \Sigma_{{\rm out}}$,
defined by gluing $\Sigma_{{\rm in, out}}$ along their boundaries in the obvious way, 
therefore has the 
same total intersection with $\Delta$, $N_+$, and $N_-$ as $\Sigma_{{\rm in}}$.
$\Sigma$
is a closed five-cycle, so that the integral
$\int d\star d C^{(5)} $ must vanish.  
The equation of motion for $C^{(5)}$ is:\foot{We follow the notation in,
for example, ref. \refs{\PeetES} in deducing factors of $g, \ell_s$ and $2\pi$.
The only difference is that we rescale $F\to (\sqrt{2\pi})\ell_s^2 F$,
so that $F$ is dimension 2 and has a standard kinetic term.}
\eqn\RReom{
	d\star d C^{(5)} = (2\pi)^3 g_s \ell_s^3
	\left( \sum_{k} \delta^{(5)}(\Sigma^{D4}_k) 
	+ \sum_i \delta^{(3)}(\Sigma^{D6}_i) \wedge F\right) \ ,
	}
where $\Sigma^{D4}_{k}$ are the set of cycles about which D4-branes are wrapped,
and $\Sigma^{D6}_i$ the set of cycles about which D6-branes are wrapped.
Integrating this equation over $\Sigma$, the result 
for our configuration is:
\eqn\intersectiontwo{
\eqalign{
	0 = & 
	\int_{\Sigma} \left(\delta^{(5)}(\Sigma^{D4}) + 
	\delta^{(3)}(N_+) \wedge F_+
	+ \delta^{(3)}(N_-) \wedge F_-\right) \cr	
	& = \Sigma \cap \Delta + 
	\left( (K\times S^1_2)\cap N_+ \right)
		\int_{\IR^2} F_+
		+ 
		\left( (K\times S^1_2)\cap N_- \right) 
		\int_{\IR^2} F_- \cr
		&  = 1 + \int_{\IR^2} (F_+ - F_-)
}}
Therefore $\int_{\IR^2} (F_+ - F_-) = -1$, 
and the string carries the magnetic flux consistent with our 
identification of this as a D-term string.

\subsubsection{Axion charge}

Next, we verify that we have identified the correct
axion charge of the string. Recall that gauge flux
$F$ on the worldvolume of D6 carries D4 charge.
Because of this, 
there is a sense in which the D4 doesn't actually end
-- the locus on which $C^{(4)}$ is sourced does not have a boundary --
but rather, half of it goes into $C_+$ and
(minus) half of it goes into $C_-$.
In this sense, there is, in the higgs phase,  actually a D4 wrapping
the {\it closed} cycle 
$ \Sigma \equiv {1\over 2} [N_+ - N_-] = {1\over 2} [C_+ - C_-
+ 2 D]$, which is
intersection-dual to $[N] = [N_+ +N_-]$.
In equations, this statement can be tested by
looking again at the equations of motion for the RR 5-form
\RReom.
Consider integrating this equation relating 5-forms
over the 5-chain $\Sigma \times B_2 $
where $B_2$ is a gaussian volume surrounding
the string in spacetime
(\ie\ $B_2$ is a 2-ball surrounding
the string in spacetime, whose boundary is an $S^1$ at infinity
encircling the string).  This gives
\eqn\gaussrollsover{ 
\int_{B_2} d \star d \int_\Sigma C^{(5)}
= (2\pi)^3 g_s \ell_s^3 \int_{B_2} 
\delta^{(2)}({\rm string}) \int_\Sigma \delta^{(3)}(D)
+ (2\pi)^3 g_s \ell_s^3 \int_{B_2} F \int_\Sigma \delta^{(3)}(N) .}
The left-hand side of this equation is $(2\pi \ell_s)^3 g_s$ because 
$ \star d \int_\Sigma C^{(5)} = da$ is the axion flux, and
$ \int_{B_2} d(da_-) = \int_{S^1_\infty} da_- = (2\pi\ell_s)^3g_s $:
the axion goes once through its period when you go around the string.\foot{The
normalization follows by demanding that equation \RReom\ is consistent
in flat space in the presence of a flat D4-brane.}
The integral on the right-hand side of \gaussrollsover\ 
is one because it is equal to the intersection product
$ \Sigma \cap N = 1 $. Indeed, this
is consistent with our claim that the D-term string is charged
under the RR axion $a_- = \int_{N_+ + N_- } C^{(3)}_{RR}$.

\newsec{String duality and D-term strings}

\lref\DabholkarYF{
A.~Dabholkar, G.~W.~Gibbons, J.~A.~Harvey and F.~Ruiz Ruiz,
``Superstrings And Solitons,''
Nucl.\ Phys.\ B {\bf 340}, 33 (1990).
}

We conclude by discussing the relation 
of our results to the
analogous physics of the heterotic string
compactified on a Calabi-Yau threefold.
In that theory, 
when the anomaly in a $U(1)$ factor
of the gauge group is cancelled by the Green-Schwarz mechanism
involving 
the NS 2-form $B$, the corresponding D-term gets a one-loop correction
proportional to the string tension $M_s^2$ \refs{\DineXK,\DineGJ,\ads}.
It is worth noting that this is 
consistent with results about 
D-term strings in \refs{\DvaliZH,\BinetruyHH}.  A D-term of order
$M_s^2$, if SUSY remains unbroken, implies a cosmic string with 
$N=(2,0)$ worldsheet supersymmetry, and tension
$M_s^2$, whose worldsheet couples linearly to $B$.  
But this is the heterotic string itself!
This is in keeping with the fact that the heterotic string
 can be written as a (singular) soliton of the massless fields of string theory \refs{\DabholkarYF}.
 
 It is also consistent with our results via string duality.  Start with the duality between
 heterotic string theory compactified on $T^3$ and M-theory compactified on $K3$.
 The heterotic string is dual to an M5-brane wrapped on the K3.
 One may fiber these dual pairs over a (large) 
three-manifold
base to find an $\CN=1$ dual pair
 of heterotic string theory compactified on a ($§T^3$-fibered) Calabi-Yau background
 and M-theory compactified on a ($K3$-fibered) $G_2$ manifold $Y$.  
This M theory compactification
can be reduced to IIA along an $S^1$-fibration of the K3 fibers of $Y$,
leading to a CY backround of IIA with 
orientifold 6-planes 
and D6-branes at the loci where the $S^1$ fiber shrinks.
The M5-brane wrapped on a K3 fiber
 becomes an open D4-brane ending on the D6-branes, 
as in section four.
This string is then a D-term string for
 one of the D6-brane gauge groups, and has tension of order $1/g_s$,
 consistent with the fact that the D-term arises at 
type IIA tree-level.

\bigskip
\centerline{\bf Acknowledgements}

We would like to thank A. Adams, O. Aharony, N. Arkani-Hamed, 
A. Dabholkar, B. Fiol, C. Hofman, S. Kachru, 
N. Saulina, H. Schnitzer, and J. Simon for useful discussions.
AL would like to thank the SLAC theory group,
the Stanford ITP, the Weizmann Institute for Science,
and the CERN Theoretical Physics Group
for their hospitality during various stages of this
project.  JM would like to thank the Aspen
Center for Physics for hospitality during the course of this work.
The research of AL is supported in part by
NSF grant PHY-0331516, and in part by the Dept. of Energy 
under Grant No. DE-FG02-92ER40706 and an Outstanding Junior Investigator award.    
The research of JM is supported by a Princeton 
University Dicke Fellowship, and by the Department of Energy under 
Grant No. DE-FG03-92ER40701.  Any opinions, findings, and conclusions or recommendations
expressed in this material are those of the authors and do not necessarily reflect the
views of the National Science Foundation.

\appendix{A}{Mode decomposition for open strings ending on intersecting D6-branes}

\def\barpsi{{\bar \psi}}

This appendix contains the explicit mode decompositions for open strings
ending on the D6-branes discussed in \S3.  The results for worldsheet fields
corresponding to the $\IR^4$ directions are standard -- the fields
are bosons satisfying Neumann boundary conditions, and their fermionic superpartners.
We will concentrate on the internal bosons and fermions.  The basic results are in 
\refs{\bdl}; 
we write them explicitly here to establish our conventions.

\subsec{Internal bosons -- closed string sector}

The closed string modes along $\IC^3$ are complex scalars
\eqn\complexmode{
	Z^i = z^i_0 + p^i \tau + {i\over \sqrt{4\pi}} 
	\sum_{n=-\infty}^{\infty} \left( {Z^i_n   \over n} e^{-i n (\tau-\sigma)} + 
		{\tilde{Z}^i_n \over n} e^{-i n(\tau + \sigma)} \right)
}
where $\tau$ runs along the time axis of the cylinder, $\sigma\in [0,2\pi]$,
$i,\bi = 1,2,3$ are complex indices, and the metric is the flat metric
$\eta_{i\jb}$ on $\IC^3$.
We will write the complex conjugate fields as
\eqn\ccmode{
	\bar{Z}^{\bi} = \bz^{\bi}_0 + \bar{p}^{\bi} \tau + 
	{i\over \sqrt{4\pi}} \sum_{n=-\infty}^{\infty} \left(
	{\bar{Z}^{\bi}_n \over n} e^{-i n (\tau - \sigma)} + 
	{\tilde{\bar{Z}}_n^{\bi}\over n} e^{-i n (\tau + \sigma)}\right)\ .
}
The canonical commutation relations imply that
\eqn\closedccr{
	[Z^i_n,\bar{Z}^{\jb}_{n'}] = n \delta_{n + n'}\eta^{i\jb}\ ;\ \ \ 
	[\tilde{Z}^i_n,\tilde{\bar{Z}}^{\jb}_{n'}] = n \delta_{n + n'}\eta^{i\jb}
}
The oscillator vacuum is defined by $Z_n |0\rangle = 0, \tilde Z_n \ket{0}=0$ for $n > 0$.

\subsec{Internal bosons -- $11$ sector}

For open strings, the worldsheet coordinate $\sigma\in[0,\pi]$.  In the $11$ sector,
the open strings satisfy the boundary conditions
\eqn\oneonebosbc{
	\Im Z^i|_{\sigma=0,\pi} = 0\ ;\ \ \ \ \Re (\p_{\sigma} Z^i)|_{\sigma =0,\pi} = 0\ .
	}
In terms of oscillator modes, this implies that $Z^i_n = - \tilde{\bar{Z}}^{\bi}_n$,
and $\Im p^i = 0$.  The Hamiltonian $L_0$ for these modes is:
\eqn\intboshamilt{
	L_0^{bos} = {1\over 4} \sum_{i} (\Re p^i)^2 + \sum_{i,\ n^i > 0} \left(\bar{Z}^{\bi}_{-n^i} 
		Z^i_{n^i} + 
		Z^i_{-n^i} \bar{Z}^{\bi}_{n^i} \right)\ .
}

\subsec{Internal bosons -- $22$ sector}

In the $22$ sector,
the open strings satisfy the boundary conditions
\eqn\twotwobosbc{
	\Im \left( e^{i\theta_i} Z^i|_{\sigma=0,\pi}\right) 
	= 0\ ;\ \ \ \ \Re \left(e^{i\theta_i} \p_{\sigma} Z^i\right)|_{\sigma =0,\pi} = 0\ .
	}
In terms of oscillator modes, this implies that $Z^i_n = - e^{-2i\theta_i}\tilde{\bar{Z}}^{\bi}_n$,
and $\Im p^i = 0$.  The Hamiltonian $L_0$ for these modes is:
\eqn\intboshamilt{
	L_0^{bos} = {1\over 4} \sum_{i}^3 (\Re p^i)^2 + \sum_{i;\ n^i > 0} \left(\bar{Z}^{\bi}_{-n^i} 
		Z^i_{n^i} + 
		Z^i_{-n^i} \bar{Z}^{\bi}_{n^i} \right)\ .
}

\subsec{Internal bosons -- $12$ and $21$ sectors}

In the $12$ sector, the strings end on brane $1$ at $\sigma = 0$ and on brane $2$ at
$\sigma = \pi$.  Therefore, 
\eqn\onetwobosbc{
\eqalign{
	\Im Z^i|_{\sigma=0} &= 0\ ;\ \ \ \ \Re (\p_{\sigma} Z^i)|_{\sigma =0} = 0\cr
	\Im \left( e^{i\theta_i} Z^i|_{\sigma=\pi}\right) 
	&= 0\ ;\ \ \ \ \Re \left(e^{i\theta_i} \p_{\sigma} Z^i\right)|_{\sigma =\pi} = 0\ .
}}
In order to satisfy these boundary conditions, we must change the moding
of the oscillators.  The mode expansion in this sector is:
\eqn\onetwomode{
\eqalign{
	Z^i &= z_0 + {i\over \sqrt{4\pi}} \sum_{n=-\infty}^{\infty}
	\left( {Z^i_{n + \alpha_i}\over n+\alpha_i} e^{-i (n+\alpha_i)(\tau - \sigma)}
	+ {\tilde{Z}^i_{n-\alpha_i}\over n-\alpha_i}
	e^{-i(n-\alpha_i)(\tau + \sigma)}\right)\cr
	\bar{Z}^{\bi} &= z_0 + {i\over \sqrt{4\pi}} \sum_{n=-\infty}^{\infty}
	\left( {\bar{Z}^{\bi}_{n + \alpha_i}\over n+\alpha_i} e^{-i (n+\alpha_i)(\tau - \sigma)}
	+ {\tilde{\bar{Z}}^{\bi}_{n-\alpha_i}\over n-\alpha_i}
	e^{-i(n-\alpha_i)(\tau + \sigma)}\right)
}}
where $\alpha_i = {\theta_i\over \pi}$.  Canonical commutation relations
imply the following commutators for the modes:
\eqn\onetwobosmodecomm{
	[Z^i_{n+\alpha_i},\bar{Z}^{\jb}_{n'+\alpha_{\jb}}] = (n+\alpha_i) \delta_{n+n'}\eta^{i\jb}\ .
}
The boundary conditions \onetwobosbc\ can then be written in terms of the modes as:
\eqn\onetwobcmodes{
	Z^i_{n+\alpha_i} = - \tilde{\bar{Z}}^{\bi}_{n+\alpha_i}\ ,
}
and the zero modes are forced to vanish, as one might expect:
motion away from this intersection forces the string to stretch out, and is not a zero mode.
The Hamiltonian for modes of $Z^i$ in this sector is:
\eqn\onetwohamgt{
\eqalign{
	L^{12,\alpha> 0}_0  & = \sum_{n \geq 0} \bar{Z}^{\bi}_{-n+\alpha} Z^i_{n+\alpha}
		+ \sum_{n>0} Z^i_{-n+\alpha_i}\bar{Z}^{\bi}_{n+\alpha}\ \ {\rm if}\ \alpha > 0\cr
	L^{12,\alpha< 0}_0  & = \sum_{n > 0} \bar{Z}^{\bi}_{-n+\alpha} Z^i_{n+\alpha}
		+ \sum_{n\geq 0} Z^i_{-n+\alpha_i}\bar{Z}^{\bi}_{n+\alpha}\ \ {\rm if}\ \alpha < 0\cr\ .
}}
Using the canonical commutation relations, the spectrum of $L_0$ in this sector is:
\eqn\onetwospectrum{
\eqalign{
   &\ E_{N,\bar{N}} = \sum_{n\geq 0} (n + \alpha) N_n + \sum_{\bar{n}>0} (
      \bar{n}-\alpha) \bar{N}_{\bar{n}}
~~{\rm if}\ \alpha > 0\cr
   &\ E_{N,\bar{N}} = \sum_{n> 0} (n + \alpha) N_n + \sum_{\bar{n}\geq 0}
      \bar{n}-\alpha) \bar{N}_{\bar{n}} 
~~{\rm if}\ \alpha < 0\cr
}}
where $N_n,\bar{N}_{\bar{n}} \in \IZ$.

\subsec{Internal fermions -- closed string sector}

The fermions $\psi_{\pm}^i,\barpsi_{\pm}^{\bi}$ are the worldsheet superpartners of $Z^i,\bar{Z}^{\bi}$.
The mode expansions in the closed string sector are:
\eqn\closedfermions{
\eqalign{
   \psi_+^i & = \sum_{n} \psi_{+,n}^i e^{-in(\tau + \sigma)}\ ;\ \ 
   \barpsi_{+}^{\bi} = \sum_{n} \barpsi_{+,n}^{\bi} e^{-in(\tau + \sigma)}\cr
   \psi_-^i & = \sum_{n} \psi_{-,n}^i e^{-in(\tau - \sigma)}\ ;\ \ 
   \barpsi_{-}^{\bi} = \sum_{n} \barpsi_{-,n}^{\bi} e^{-in(\tau - \sigma)}\cr
}}
The $U(1)_R$ currents are $J_{R,\pm} = \eta_{i\jb}\psi^i_{\pm}\barpsi^{\jb}_{\pm}$.
Fermions $\psi^i_{\pm}$ have charge $+1$ under this $U(1)_R$, while
the fermions $\barpsi^{\jb}_{\pm}$ have charge $-1$.
For NS fermions, $n\in \IZ + \half$, while for R fermions, $n\in \IZ$.
The canonical anticommutation relations for $\psi(\sigma)$ imply the following
anticommutation relations for the modes:
\eqn\csfermionmodecomm{
   \{\psi_{\pm,n}^i,\barpsi_{\pm,n'}^{\jb}\} = \delta_{n+n'}\eta^{i\jb}
}

The vacuum is defined by $\psi^i_{\pm,n}|0\rangle = 0$ for $n > 0$.  
For R fermions, the $n=0$ modes form a Clifford algebra: 
\eqn\clifford{
   \{\psi^i_{\pm,0},\barpsi^{\jb}_{\pm,0}\} = \eta^{i\jb}
}
For a given complex direction $i$, 
this algebra has a two-dimensional representation.
For example, for $\psi^1_{0,+},\barpsi^{\bar{1}}_{0,+}$, we can write
\eqn\groundrep{
\eqalign{
   & \barpsi^{\bar{1}}_{0,+} |\downarrow_{1,+}\rangle = 0\cr
   & \psi^1_{0,+}|\downarrow_{1,+}\rangle = |\uparrow_{1,+}\rangle\cr
   & \psi^1_{0,+}|\uparrow_{1,+}\rangle = 0
}}
CPT invariance requires that $|\uparrow_{i,\pm}\rangle$ and $|\downarrow_{i,\pm}\rangle$
have R-charges $+\half$ and $-\half$, respectively.

\subsec{Internal fermions -- $11$ sector}

The boundary conditions for the fermions are related via supersymmetry to
those of the bosons.  In the $11$ sector, these conditions are:
\eqn\oneonefermionbc{
   \psi^i_{+} = \barpsi_-^{\bi}\ .
}
at both $\sigma = 0,\pi$.
In terms of modes, this implies:
\eqn\oneonefmodebc{
   \psi^i_{+,n} = \barpsi^{\bi}_{-,n}
}
The worldsheet Hamiltonian for these modes is:
\eqn\oneonefermham{
   L_0 = \sum_{n>0} n \left( \psi^{\bi}_{+,-n} \psi^i_{+,n} + \psi^i_{+,-n} \psi^{\bi}_{+,n}\right)
}

\subsec{Internal fermions -- $22$ sector}

In the $22$ sector, the boundary conditions are:
\eqn\oneonefermionbc{
   \psi^i_{+} =e^{-2i\theta_i} \barpsi_-^{\bi}\ .
}
at both $\sigma = 0,\pi$. In terms of modes, this implies:
\eqn\oneonefmodebc{
   \psi^i_{+,n} = e^{-2i\theta_i}\barpsi^{\bi}_{-,n}
}
The worldsheet Hamiltonian is the same as \oneonefermham.

\subsec{Internal fermions -- $12$ and $21$ sectors}

Here the boundary conditions are
\eqn\onetwofermbc{
\eqalign{
   \psi_+^i (\sigma = 0) &= \barpsi_-^{\bi} (\sigma=0)\cr
   \psi_+^i (\sigma = \pi) &= e^{-2i\theta_i}\barpsi_-^{\bi} (\sigma=\pi)
}}
To solve this, we must shift the moding:
\eqn\onetwofermmodes{
\eqalign{
   \psi^i_{\pm} & = \sum_n \psi^i_{\pm, n\pm \alpha_i} e^{-i(n\pm\alpha_i)(\tau \pm \sigma)}\cr
   \barpsi^{\bi}_{\pm}&=\sum_n\barpsi^{\bi}_{\pm, n\pm \alpha_i} e^{-i(n\pm\alpha_i)(\tau \pm \sigma)}
}}
where $n\in\IZ+\half$ for NS fermions, and $n\in\IZ$ for R fermions.  The 
anticommutation relations for the modes are:
\eqn\onetwofermcomm{
   \{\psi^i_{\pm,n\pm\alpha_i},\barpsi_{\pm,n'\pm\alpha_i}^{\jb}\} = \eta^{i\jb}\delta_{n+n'}\ ,
}
and the boundary conditions are:
\eqn\onetwofermbc{
\eqalign{   
   & \psi^i_{+,n+\alpha_i} = \barpsi^{\bi}_{-,n-\alpha_i}\cr
   & \barpsi^{\bi}_{+,n+\alpha_i}= \psi^i_{-,n-\alpha_i}
}}

The Hamiltonian in this sector is:
\eqn\onetwoham{
   L_0 = \sum_{n > -\alpha_i} (n+\alpha_i) \barpsi^{\bi}_{+,-n+\alpha_i}\psi^i_{+,n+\alpha_i}
   + \sum_{n> \alpha_i} (n-\alpha_i) \psi^i_{-,-n+\alpha_i}\barpsi^{\bi}_{+,n+\alpha_i}
}
with the spectrum
\eqn\onetwofermspectrum{
   E_{N,\bar{N}} = \sum_{n>-\alpha_i} (n+\alpha_i)N_n + \sum_{n>\alpha_i} (n-\alpha_i)\bar{N}_{\bar{n}}
}
Here $N,\bar{N} = 0$ or $1$.

\listrefs
\end